\documentclass[twocolumn]{aastex631}
\usepackage{amsmath}
\usepackage{mathtools}

\newcommand{\Gyr}{\,{\rm Gyr}}
\newcommand{\mgfe}[0]{[{\rm Mg/Fe}]}
\newcommand{\ofe}[0]{[{\rm O/Fe}]}
\newcommand{\sife}[0]{[{\rm Si/Fe}]}

\newcommand{\afe}[0]{[\alpha/{\rm Fe}]}

\newcommand{\mgh}{[{\rm Mg}/{\rm H}]}
\newcommand{\feh}[0]{[{\rm Fe/H}]}

\newcommand{\xfe}{[{\rm X}/{\rm Fe}]} 
\newcommand{\xfecc}{\xfe_{\rm cc}}
\newcommand{\mgfecc}{[{\rm Mg}/{\rm Fe}]_{\rm cc}}
\newcommand{\ofecc}{[{\rm O}/{\rm Fe}]_{\rm cc}}
\newcommand{\sifecc}{[{\rm Si}/{\rm Fe}]_{\rm cc}}
\newcommand{\afecc}{[\alpha/{\rm Fe}]_{\rm cc}}
\newcommand{\afeeq}{[\alpha/{\rm Fe}]_{\rm eq}}

\newcommand{\logg}{\log(g)}
\newcommand{\teff}{T_{\rm eff}}
\newcommand{\kpc}{\rm \; kpc}

\newcommand{\Msun}{M_{\odot}}
\newcommand{\mxexp}{m_{X,{\rm exp}}}
\newcommand{\mxwind}{m_{X,{\rm wind}}}

\newcommand{\yxcc}{y_{\rm X}^{\rm cc}}
\newcommand\ybarxcc{\bar{y}_{\rm X}^{\rm cc}}
\newcommand{\yfecc}{y_{\rm Fe}^{\rm cc}}
\newcommand{\ybarfecc}{\bar{y}_{\rm Fe}^{\rm cc}}
\newcommand{\yocc}{y_{\rm O}^{\rm cc}}
\newcommand\ybarocc{\bar{y}_{\rm O}^{\rm cc}}
\newcommand{\ymgcc}{y_{\rm Mg}^{\rm cc}}
\newcommand\ybarmgcc{\bar{y}_{\rm Mg}^{\rm cc}}
\newcommand{\ysicc}{y_{\rm Si}^{\rm cc}}
\newcommand\ybarsicc{\bar{y}_{\rm Si}^{\rm cc}}
\newcommand{\yfeIa}{y_{\rm Fe}^{\rm Ia}}
\newcommand{\ybarfeIa}{\bar{y}_{\rm Fe}^{\rm Ia}}

\newcommand{\Zxsun}{Z_{{\rm X},\odot}}
\newcommand{\Zfesun}{Z_{{\rm Fe},\odot}}
\newcommand{\Zosun}{Z_{{\rm O},\odot}}
\newcommand{\Zmgsun}{Z_{{\rm Mg},\odot}}
\newcommand{\Zsisun}{Z_{{\rm Si},\odot}}
\newcommand{\Zxeq}{Z_{\rm X,eq}}
\newcommand{\Xd}{X_{\rm D}}
\newcommand{\Xdp}{X_{\rm D}^{\rm P}}

\newcommand{\taustar}{\tau_*}
\newcommand{\tausfh}{\tau_{\rm SFH}}
\newcommand{\taubar}{\bar{\tau}}
\newcommand{\tauIa}{\tau_{\rm Ia}}

\newcommand{\Fexp}{F_{\rm exp}}
\newcommand{\Fimf}{F_{\rm IMF}}
\newcommand{\Nmassive}{N_{\rm massive}}
\newcommand{\Ncc}{N_{\rm cc}}
\newcommand{\Rcc}{R_{\rm cc}}
\newcommand{\RIa}{R_{\rm Ia}}
\newcommand{\Msn}{M_{\rm SN}}
\newcommand{\Mmin}{M_{\rm min}}
\newcommand{\Mmax}{M_{\rm max}}
\newcommand{\Mrem}{M_{\rm rem}}
\newcommand{\Mdotstar}{\dot{M}_*}
\newcommand{\Mdotout}{\dot{M}_{\rm out}}
\newcommand{\etasun}{\eta_\odot}
\newcommand{\Rgal}{R_{\rm gal}}

\newcommand{\Eqref}[1]{Equation~(\ref{eqn:#1})}

\shorttitle{Implications of the CCSN Fe-yield}
\shortauthors{Weinberg et al.}

\graphicspath{{./}{figures/}}

\begin{document}

\title{The Scale of Stellar Yields: Implications of the Measured Mean Iron Yield of Core Collapse Supernovae}

\correspondingauthor{David Weinberg}
\email{weinberg.21@osu.edu}

\author[0000-0001-7775-7261]{David H. Weinberg}
\affiliation{The Department of Astronomy and Center of Cosmology and AstroParticle Physics, The Ohio State University, Columbus, OH 43210, USA}

\author[0000-0001-9345-9977]{Emily J. Griffith}
\altaffiliation{NSF Astronomy and Astrophysics Postdoctoral Fellow}
\affiliation{Center for Astrophysics and Space Astronomy, Department of Astrophysical and Planetary Sciences, University of Colorado, 389 UCB, Boulder, CO 80309-0389, USA}

\author[0000-0002-6534-8783]{James W. Johnson}
\affiliation{The Department of Astronomy and Center of Cosmology and AstroParticle Physics, The Ohio State University, Columbus, OH 43210, USA}
\affiliation{The Observatories of the Carnegie Institution for Science, 813 Santa Barbara St., Pasadena, CA, 91101}

\author[0000-0003-2377-9574]{Todd A. Thompson}
\affiliation{The Department of Astronomy and Center of Cosmology and AstroParticle Physics, The Ohio State University, Columbus, OH 43210, USA}

\begin{abstract}
The scale of $\alpha$-element yields is difficult to predict from theory because of uncertainties in massive star evolution, supernova physics, and black hole formation, and it is difficult to constrain empirically because the impact of higher yields can be compensated by greater metal loss in galactic winds.  We use a recent measurement of the mean iron yield of core collapse supernovae (CCSN) by Rodriguez et al.\ (RMN23), $\ybarfecc=0.058 \pm 0.007 \Msun$, to infer the scale of $\alpha$-element yields by assuming that the plateau of $\afe$ abundance ratios observed in low metallicity stars represents the yield ratio of CCSN.  For a Kroupa (2001) initial mass function and a plateau at $\afecc=0.45$, we find that the population-averaged yields of O and Mg are about equal to the solar abundance of these elements, $\log\yocc/\Zosun = \log \ymgcc/\Zmgsun = -0.01 \pm 0.1$, where $\yxcc$ is the mass of element X produced by massive stars per unit mass of star formation.  The inferred O and Fe yields agree with predictions of the Sukhbold et al.\ (2016) CCSN models assuming their Z9.6+N20 neutrino-driven engine, a scenario in which many progenitors with $M<40 M_\odot$ implode to black holes rather than exploding.  The yields are lower than assumed in some models of galactic chemical evolution (GCE) and the galaxy mass-metallicity relation, reducing the level of outflows needed to match observed abundances.  For straightforward assumptions, we find that one-zone GCE models with $\eta=\Mdotout/\Mdotstar \approx 0.6$ evolve to solar metallicity at late times.  The ISM D/H ratio predicted by these models is about 70\% of the primordial D/H ratio, which is lower than observational estimates but consistent at $2\sigma$.  By requiring that models reach $\afe \approx 0$ at late times, and assuming a mean Fe yield of $0.7\Msun$ per Type Ia supernova, we infer a Hubble-time integrated SNIa rate of $1.1 \times 10^{-3} \Msun^{-1}$, compatible with estimates from supernova surveys.  The RMN23 measurement provides one of the few empirical anchors for the absolute scale of nucleosynthetic yields, with wide-ranging implications for stellar and galactic astrophysics.
\end{abstract}


\section{Introduction} \label{sec:intro}

The nucleosynthetic yields of elements are the most basic ingredient of galactic chemical evolution (GCE) models because they determine the rate at which stars enrich their surroundings.  The relative yields of different elements can be constrained empirically through abundance ratios in stellar populations, using time delays and theoretical models as a guide to separate the contributions of different sources such as core collapse supernovae (CCSN), Type Ia supernovae (SNIa), and asymptotic giant branch (AGB) stars (e.g., \citealt{Griffith2019,Griffith2022galah,Weinberg2019,Weinberg2022,Johnson2023}).  However, the absolute scale of yields is difficult to constrain empirically because it is largely degenerate with the effect of outflows (a.k.a.\ galactic winds), which remove newly produced metals from the star-forming interstellar medium (ISM).  It is widely accepted that the mass-metallicity relation \citep{Tremonti2004,Andrews2013} is primarily driven by an increasing efficiency of outflows in the shallower potential wells of lower mass galaxies \citep{Finlator2008,Peeples2011,Dave2012,Zahid2012,Lin2022}.  For the Milky Way, on the other hand, some GCE models assume substantial metal outflows (e.g., \citealt{Schoenrich2009,Johnson2021}) while others do not (e.g., \citealt{Minchev2013,Spitoni2019}), with both classes reproducing observed chemical abundances because they adopt significantly different population-averaged yields.\footnote{We use the term population-averaged to encompass averaging over the stellar initial mass function (IMF) and other properties such as rotation and binary fractions that may affect yields.}  The absolute scale of yields is also crucial for assessing the heavy element budget of galaxies and predicting the abundance of metals in the circumglactic medium (e.g., \citealt{Peeples2014}).

Predicting nucleosynthetic yields of CCSN is a long-standing goal of supernova modeling (e.g., \citealt{Woosley1995,Chieffi2013,Nomoto2013,Sukhbold2016,Limongi2018,Curtis2021}).  Given yields as a function of progenitor mass and metallicity, there are two further challenges in calculating the population-averaged yield.  The first is the choice of IMF; for example, models such as \cite{Minchev2013} and \cite{Spitoni2019} have a low population-averaged $\alpha$-element yield because they adopt a steep IMF \citep{Scalo1986,Kroupa1993}, while models such as \cite{Schonrich2009} and \cite{Johnson2021} adopt a \cite{Kroupa2001} IMF with larger numbers of high mass stars.  \cite{Vincenzo2016yield} demonstrate the large impact that the choice of IMF can have on population-averaged yields.  The second and perhaps even thornier challenge is the uncertain physics of black hole formation, because in many cases the massive star progenitors that are most efficient in producing specific elements are also those most susceptible to forming black holes and releasing no heavy elements at all (e.g., \citealt{Sukhbold2016}, hereafter S16).  Griffith et al. (\citeyear{Griffith2021snyield}, hereafter G21) show that plausible variations in the degree of black hole formation can produce a factor of three variation in the population-averaged CCSN yield of O and Mg.  

In a recent study, Rodriguez et al.\ (\citeyear{Rodriguez2022}, hereafter RMN23) estimate the mean Fe yield of stripped envelope CCSN (Type Ib, Ic), which they combine with a similar estimate for Type II CCSN \citep{Rodriguez2021} and the relative frequency of CCSN types \citep{Shivvers2017} to infer the mean Fe yield of CCSN, $\ybarfecc = 0.058 \pm 0.007 \Msun$.  The Fe yield of individual supernovae can be estimated from the late-time light curve, which is powered by the radioactive decay of $^{56}$Ni to $^{56}$Co, leading after a further decay to $^{56}$Fe.  Here we explore the implications of the RMN23 yield determination for other population-averaged yields, for galactic outflows, for black hole formation, and for the time-integrated rate of SNIa.  In addition to Fe, we focus on O and Mg, which are both $\alpha$-elements thought to come almost entirely from CCSN \citep{Andrews2017,Rybizki2017,Johnson2019}, and on the $\alpha$-element Si, which has a sub-dominant but not negligible contribution from SNIa \citep{Griffith2022galah,Weinberg2022}.  

Our basic assumption --- standard in GCE modeling --- is that the plateau observed in [O/Fe], [Mg/Fe], and [Si/Fe] at low metallicity reflects the population-averaged yield ratios of CCSN, with the decline at higher metallicity caused by the SNIa contribution to Fe.  Combined with the RMN23 value of $\ybarfecc$, the observed plateau level allows us to infer the mean yield per supernova of O, Mg, and Si, which we denote $\ybarocc$, $\ybarmgcc$, $\ybarsicc$.  What matters for GCE models is the yield per unit mass of stars formed, which we denote with $y$ instead of $\bar{y}$.  Going from $\bar{y}$ (with units of $\Msun$) to $y$ (dimensionless) requires a choice of IMF and the fraction $\Fexp$ of massive stars ($M>8\Msun$) that explode as CCSN.  The theoretical uncertainty in $\Fexp$ is significant (see \S\ref{sec:fexp}), but it is much smaller than the uncertainty in the yield because the {\it number} of massive stars is dominated by those of relatively low mass that are relatively easy to explode.

Our translations from $\yxcc$ values to outflow constraints rely on the analytic GCE models of Weinberg et al. (\citeyear{Weinberg2017equ}, hereafter WAF), with arguments similar to those made in the mass-metallicity context by \cite{Peeples2011}, \cite{Dave2012}, \cite{Zahid2012}, and \cite{Lin2022}, and in the dwarf galaxy context by \cite{Johnson2022dwarf} and \cite{Sandford2022}.  The drop between the plateau values of [X/Fe] and the solar ratios, $\xfe \approx 0$, characteristic of the present-day local disk depends mainly on the population-averaged Fe yield of SNIa relative to that of CCSN.  Because the mean Fe yield per SNIa is reasonably well established at $\ybarfeIa=0.6-0.7\Msun$ (e.g., \citealt{Howell2009}), $\ybarfecc$ can be used to normalize the SNIa delay time distribution (DTD) and hence the number of SNIa produced per unit mass of star formation.  We also examine what the inferred Fe, O, Mg, and Si yields imply for the CCSN models of S16/G21, in particular about which massive stars explode.

Throughout this paper we assume a \cite{Kroupa2001} IMF, with $dN/dM \propto M^{-2.3}$ for $M=0.5-120\Msun$ and $dN/dM \propto M^{-1.3}$ for $M=0.08-0.5\Msun$.  Although the functional form is different, this IMF is similar to that of \cite{Chabrier2003}, with both having a similar high mass slope but fewer low mass stars than a \cite{Salpeter1955} IMF ($dN/dM \propto M^{-2.35}$).  We refer to our choice as simply a Kroupa IMF, but we caution that it is quite different from that of \cite{Kroupa1993}, which has a $-2.7$ high mass slope and thus leads to much smaller predicted yields (\citealt{Vincenzo2016yield}; G21).  

In Section~\ref{sec:inference} we combine the RMN23 value of $\ybarfecc$ with the estimated level of the $\afe$ plateau to infer the overall scale of CCSN element yields.  In Section~\ref{sec:implications} we examine the implications of this inferred scale for galactic outflows, for the SNIa rate, for theoretical models of CCSN and black hole formation, and for the deuterium abundance of the ISM.  The logical flow of these arguments is summarized in Figure~\ref{fig:summary} (Section~\ref{sec:overview}).  In the remainder of Section~\ref{sec:discussion} we discuss sources of uncertainty in our results and broader implications for the mass-metallicity relation and the chemical evolution of the Milky Way.  Section~\ref{sec:conclusions} summarizes our conclusions.

\section{Inferring the scale of yields}
\label{sec:inference}

\subsection{The $\afe$ plateau}
\label{sec:plateau}

In our analysis below we consider O, Mg, and Si as representative $\alpha$-elements, using all three because each is affected by different observational and theoretical uncertainties.  Our goal is to combine the RMN23 estimate of $\ybarfecc$ with empirical estimates of $\xfecc$ for these elements to infer their IMF-averaged CCSN yields, where $\xfecc$ represents the element ratio that would be produced by CCSN alone.  In most observational studies, the values of [O/Fe], [Mg/Fe], and [Si/Fe] show an approximately flat trend (with significant scatter) in Milky Way halo stars with $-2 \leq \feh \leq -1$. 
This plateau is usually taken to reflect the yield ratio of CCSN, on the assumption that SNIa have not yet contributed significantly to the Fe abundances of these stars.  However, the value of the plateau varies noticeably from study to study.  For the most part these differences reflect the systematic uncertainties in determining the absolute abundances from observed stellar spectra, which are more acute in low metallicity stars because they are further from solar calibration and may be more susceptible to the impact of departures from local thermodynamic equilibrium on spectral synthesis.  

The pioneering study of halo populations by \cite{Nissen2010} exhibits plateaus at $\mgfe \approx 0.35$ and $\sife \approx 0.3$ for stars they identify with the {\it in situ} halo.  \cite{Bensby2017} show trends from microlensed bulge stars and the solar neighborhood sample of \cite{Bensby2014}, exhibiting a similar Si plateau and a slightly higher Mg plateau at $\mgfe \approx 0.4$.  \cite{Kobayashi2020} present compilations of $\xfe$ measurements from many data sets.  For Mg, the \cite{Zhao2016} and \cite{Reggiani2017} data sets imply a plateau at $\mgfe \approx 0.3$, but the \cite{Andrievsky2010} data imply a higher $\mgfe \approx 0.6$ for very low metallicity stars with $-3.5 \leq \feh \leq -2.5$.  For Si, the \cite{Zhao2016} data set implies a plateau at $\sife \approx 0.3$, while the values from \cite{Cayrel2004} and \cite{Honda2004} are $\sim 0.2$ dex higher.  For O, the \cite{Zhao2016} and \cite{Amarsi2019} data sets imply a plateau at $\ofe \approx 0.6$, though it is not perfectly flat.

\begin{figure*}[!htb]
    \centering
    \includegraphics[width=\textwidth]{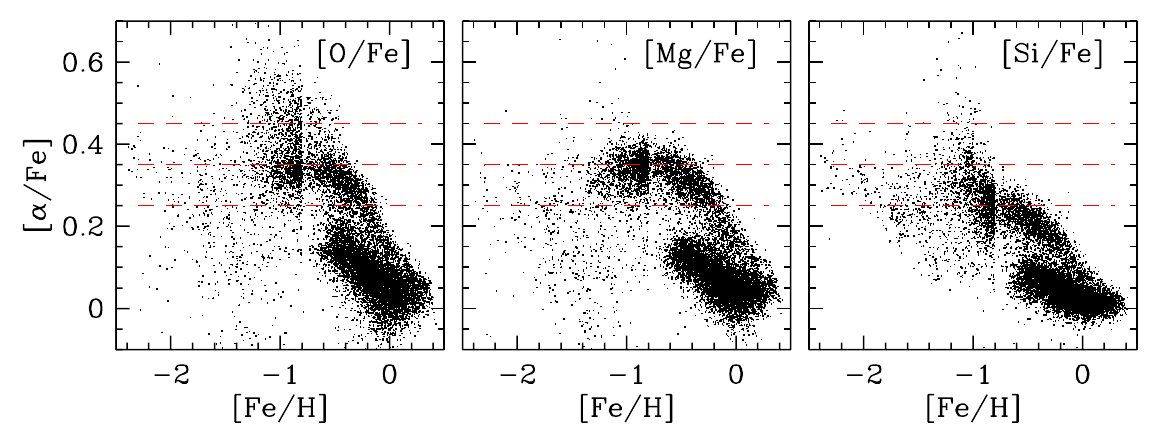}
    \caption{Distribution of APOGEE disk and halo stars in $\afe$-$\feh$ for the $\alpha$-elements O (left), Mg (middle), or Si (right).  Above $\feh=-0.8$ stars are randomly downsampled by a factor of 10; the transition to full sampling produces the edge at $\feh=-0.8$.  Horizontal lines show $\afe = 0.25$, 0.35, 0.45 for visual reference.}
    \label{fig:apogee}
\end{figure*}

Figure~\ref{fig:apogee} shows [O/Fe], [Mg/Fe], and [Si/Fe] for stars from APOGEE \citep{Majewski2017} as reported in SDSS Data Release 17 (DR17; \citealt{DR17}).  We selected stars with $0.5 < \logg < 3.5$, $3800 < \teff < 5200$, signal-to-noise ratio ${\rm SNR} > 200$ per pixel, Galactocentric radius $\Rgal<15\kpc$, and midplane distance $|Z|<10\kpc$, using distances from the DR17 AstroNN catalog \citep{Leung2019a}.  As a rough cut to reject accreted halo stars, we require ${\rm [Al/Fe]}>-0.05$ \citep{Hawkins2015,Belokurov2022}.  Changes to these cuts do not change the qualitative appearance of the plot, which shows all selected stars with $\feh < -0.8$ and a random 10\% selection at higher $\feh$.  The high-$\alpha$ stars with $-1.2 < \feh < -0.6$ suggest plateaus at approximately $+0.35$ for O and Mg and $+0.25$ for Si.  However, at lower metallicity the scatter is large and there is no clear plateau.  Furthermore, motivated by data from the H3 survey, \cite{Conroy2022} present a model in which the true CCSN plateau of the {\it in situ} halo lies at $\mgfe \approx 0.6$ but is seen only at $\feh \la -2.5$, and the trend between $-2.5 < \feh < -0.5$ is governed by a rapidly accelerating star formation efficiency that maintains a roughly constant SNIa/CCSN ratio (see Figure~\ref{fig:conroy} below).  \cite{Maoz2017}, using chemical evolution models motivated by the cosmic star formation history, also argue that $\mgfe\approx 0.3$ is an intermediate plateau arising from balanced CCSN and SNIa enrichment rather than reflecting the CCSN ratio.

Since plausible observational values for the $\afe$ plateau range from 0.3 to 0.6, we have decided to take 0.45 as a fiducial value for $\ofecc$ and $\mgfecc$.  The 0.15-dex systematic uncertainty in this choice is, unfortunately, a large (40\%) source of uncertainty in our eventual conclusions about the scale of yields.  Based on the separation between the [Si/Mg] ratios of the low-$\alpha$ and high-$\alpha$ disk populations, \cite{Weinberg2022} infer that a fraction 0.81 of the solar Si abundance arises from CCSN, with the remainder from SNIa.  We therefore adopt $0.45+\log(0.81) = 0.36$ as our fiducial value for $\sifecc$.  This 0.09-dex difference is consistent with observed differences between the [Si/Fe] and [Mg/Fe] plateaus in Figure~\ref{fig:apogee} and in the observational studies discussed previously.  We sometimes use the notation $\afe$ to refer generically to either $\ofe$ or $\mgfe$, or conceptually to an $\alpha$-element that is produced entirely by CCSN.

Because observed abundance ratios are scaled to solar values, our choice of solar abundances also matters for our results.  We adopt solar photospheric abundances from \cite{Magg2022}, as this study uses state-of-the-art atmospheric models and finds consistency between spectroscopically derived abundances and helioseismic models.  Relative to the photospheric abundances of \cite{Asplund2009}, these abundances are 0.08 dex higher for O and Si, 0.05 dex lower for Mg, and the same for Fe.

Table~5 of \cite{Magg2022} reports solar photospheric abundances on the conventional scale $x = 12+\log ({\rm X}/{\rm H})$ of 8.77, 7.55, 7.59, and 7.50 for O, Mg, Si, and Fe, respectively.  We add 0.04 dex to obtain proto-solar abundances corrected for the impact of diffusion and gravitational settling \citep{Turcotte1998}, the same correction adopted by \cite{Asplund2009}.  Our calculations require a mass fraction $Z_X$ of these elements, which we compute as
\begin{equation}
    \log Z_{\rm X} = (x-12) + \log 0.71 + \log A ~,
    \label{eqn:Zx}
\end{equation}
where 0.71 is the assumed solar hydrogen mass fraction and $A$ is the mean atomic weight, which we take to be 16.0, 24.3, 28.09, and 55.85 for O, Mg, Si, Fe.  We therefore adopt
\begin{multline}
    \{\Zosun,\Zmgsun,\Zsisun,\Zfesun\}= \\
    \{73.3,6.71,8.51,13.7\} 
    \times 10^{-4} ~.
    \label{eqn:Zsolar}
\end{multline}

\subsection{The explosion fraction}
\label{sec:fexp}

Parametric theoretical studies of the supernova mechanism tuned to produce SN 1987A generate a complicated landscape of successful explosions and unsuccessful collapses that form black holes \citep{Ugliano2012,Pejcha2015,Sukhbold2016,Ebinger2019}. Such models typically produce explosion fractions of 60-90\% (see, e.g., Figure~\ref{fig:eofm} below).  In general, lower mass progenitors explode relatively easily, but they do so with low Ni yields, low energies, and low ejected O masses. Although information on the metallicity dependence of $\Fexp$ is limited, the \cite{Pejcha2015} and \cite{Ebinger2020} models show that black hole formation should increase (and $\Fexp$ decrease) at lower metallicity. 

On the observational side, \cite{Horiuchi2011} compare the number of observed supernovae to that expected from the star formation rate, finding a discrepancy at the factor of two level that indicates either $\sim50$\,\% black hole formation fraction or an undercounting of underluminous (and underenergetic) supernovae in surveys. From the observation of a single disappearing massive star and a sample of successful supernovae, Neustadt et al. (\citeyear{Neustadt2021}, continuing the program of \citealt{Kochanek2008,Gerke2015,Adams2017}) report a black hole formation fraction $F_{\rm bh}=1-\Fexp$ of $\sim20-40$\,\%.  We will adopt $\Fexp=0.75$ as our fiducial value in calculations below, but most of the range $0.5 < \Fexp < 1$ is possible, and the value may be metallicity dependent.

\subsection{The $\alpha$-element yield}
\label{sec:yields}

By definition,
\begin{equation}
    \xfecc = \log\left({\ybarxcc/\ybarfecc \over \Zxsun/\Zfesun} \right)~.
    \label{eqn:xfepl}
\end{equation}
We can rearrange this definition to find
\begin{align}
    \ybarxcc =& \,\ybarfecc \cdot 10^{\xfecc} \cdot {\Zxsun\over\Zfesun} \label{eqn:ybarxcc1} \\
             =& \,0.163 \left({\Zxsun \over \Zfesun}\right)\left({\ybarfecc \over 0.058\Msun}\right) \nonumber \\
             & \times 10^{({\xfecc - 0.45})} \,\Msun~. \label{eqn:ybarxcc2}
\end{align}
Based on the discussion in \S\ref{sec:plateau}, we adopt
$\ofecc=\mgfecc=0.45$, $\sifecc=0.36$, and the \cite{Magg2022} solar abundances to obtain
\begin{multline}
    \{\ybarocc,\ybarmgcc,\ybarsicc,\ybarfecc\} = \\
    \{0.87,0.080,0.082,0.058\}\Msun~
    \label{eqn:ybarcc}
\end{multline}
for the mean yield per CCSN.

For purposes of GCE modeling, we want to specify the population-averaged yield in terms of $\yxcc$, mass of element X produced per unit mass of stars formed.  For a Kroupa IMF, the number of massive stars per unit mass of star formation is
\begin{equation}
    {\Nmassive \over M_*} = 
    \frac{ \int_{\Msn}^{\Mmax} {dN \over dM} dM} 
         {\int_{\Mmin}^{\Mmax} M {dN \over dM} dM } = 0.0109 \Msun^{-1},
\label{eqn:nmassive}
\end{equation}
where the numerical value assumes mass limits $\Mmin=0.08\Msun$ and $\Mmax=120\Msun$ and a threshold mass $\Msn=8\Msun$ for producing a CCSN.  For a \cite{Salpeter1955} IMF the ratio drops to 0.0068 because of the larger number of low mass stars.  For a Kroupa IMF with a higher supernova threshold $\Msn=11\Msun$ it drops to 0.0071; for $\Msn=8\Msun$ and $\Mmax=60\Msun$ it is 0.0104.  To compute the number of supernovae we also need to know the fraction $\Fexp$ of massive stars that explode as CCSN rather than collapsing to black holes.  We thus have a core collapse supernova ratio
\begin{equation}
    \Rcc \equiv \frac{\Ncc}{M_*} = 0.0109 \Fexp\Fimf \Msun^{-1}~,
    \label{eqn:ncc}
\end{equation}
where the $\Fimf$ factor corrects for departures from a Kroupa IMF.  The $\Fexp$ factor is $\leq 1$ by definition, while the $\Fimf$ factor may be $<1$ (e.g., for \citealt{Salpeter1955} or \citealt{Kroupa1993}) or $>1$ for a ``top heavy'' IMF.
Based on the discussion in \S\ref{sec:fexp} and in \S\ref{sec:s16} below, we adopt $\Fexp=0.75$ and $\Fimf=1$ as fiducial values.

The mean yield per CCSN can be converted to a mean yield per unit mass of star formation with
\begin{equation}
    \yxcc = \Rcc\ybarxcc ~.
    \label{eqn:yxcc}
\end{equation}
Using \Eqref{ybarxcc2}\ and \Eqref{ncc}\ gives
\begin{align}
    {\yxcc \over \Zxsun} =& 0.973 \left({\ybarfecc \over 0.058 \Msun}\right) 
                                 \left({\Fexp \over 0.75}\right) \left({\Fimf \over 1.0}\right) \nonumber \\
                          & \times \left({0.00137 \over \Zfesun}\right) \times 10^{(\xfecc-0.45)}~.
    \label{eqn:yxccscale}
\end{align}
Because $\xfecc$ is expressed relative to the solar abundance ratio, this formula for $\yxcc$ in solar units does not depend on the adopted solar abundance of element X.  For Si we infer a lower value of $\yxcc/\Zxsun$ because we adopt $\sifecc = 0.36$, which corresponds to $\approx 20\%$ of solar Si arising from SNIa instead of CCSN.

\Eqref{yxccscale}\ is our first key result.  Although the values of $\Fexp$, $\Fimf$, and $\xfecc$ are model dependent, and the values of $\ybarfecc$ and $\Zfesun$ have observational uncertainties, the equation itself follows directly from the definition of these quantities, independent of supernova or GCE models.  For our fiducial parameter values, the RMN23 measurement of $\ybarfecc$ implies that the population-averaged CCSN yields of $\alpha$ elements are about equal to the solar abundance of those elements, a finding that will have important implications for galactic outflows in chemical evolution models (Section \ref{sec:outflows}).  This empirical conclusion does not rely on models of massive stars and CCSN except through the choice of $\Fexp$.

Adopting the fiducial scalings of \Eqref{yxccscale}, the $\xfecc$ values of Section \ref{sec:plateau}, and the solar abundances of \Eqref{Zsolar}, we obtain
\begin{multline}
    \{\yocc,\ymgcc,\ysicc,\yfecc\} = \\
    \{71.3,6.52,6.72,4.73\} \times 10^{-4}
    \label{eqn:yxccvals}
\end{multline}
for the population-averaged CCSN yields.  Our calculations assume that these yields are independent of metallicity.  We discuss uncertainties associated with this assumption in Section~\ref{sec:uncertainties} below.

\section{Implications of the inferred yield scale}
\label{sec:implications}

\subsection{Implications for outflows}
\label{sec:outflows}

In the one-zone GCE models described by WAF, the ISM mass fraction of an element X produced by CCSN with a metallicity-independent yield evolves to an equilibrium abundance
\begin{equation}
    \Zxeq  = {\yxcc \over 1+\eta-r-\taustar/\tausfh}~.
    \label{eqn:Zeq}
\end{equation}
Here $\eta \equiv \Mdotout/\Mdotstar$ is the mass-loading factor of outflows, $r$ is the recycling factor ($\approx 0.4$ for a Kroupa IMF), $\taustar = M_{\rm gas}/\Mdotstar$ is the star formation efficiency (SFE) timescale, and the star formation history (SFH) is assumed to be exponential with $\Mdotstar \propto e^{-t/\tausfh}$.  The analytic solution approximates the recycling of material from stellar envelopes as instantaneous, returning gas at the birth metallicity to the ISM at a rate $r\Mdotstar$, which (as shown by WAF) gives chemical evolution tracks nearly identical to those of a numerical calculation with time-dependent recycling.  The full time evolution for this SFH is
\begin{equation}
Z_{\rm X}(t) = \Zxeq\left(1-e^{-t/\taubar}\right)~,
\label{eqn:Zevol}
\end{equation}
where 
\begin{equation}
\taubar = \dfrac{\taustar}{(1+\eta-r-\taustar/\tausfh)}~.  
\label{eqn:taubar}
\end{equation}
For a linear-exponential SFH, $\Mdotstar \propto t e^{-t/\tausfh}$, the late-time equilibrium is the same but the evolution to that equilibrium is slower (WAF, Equation~56).  

Observations of gas phase and Cepheid abundances imply that the ISM metallicity is approximately solar at $\Rgal=8\kpc$ in the present-day Galaxy (e.g., \citealt{Lemasle2013,daSilva2022,Esteban2022}). 
We can invert \Eqref{Zeq}\ to find the value of $\eta$ that is required to produce a solar metallicity ISM at equilibrium,
\begin{equation}
    \eta_\odot = \yxcc/\Zxsun - 1 + r + \taustar/\tausfh~.
    \label{eqn:etasun}
\end{equation}
\cite{Johnson2021} present detailed GCE models of the Milky Way disk that reproduce a wide range of observational constraints.  They base $\tausfh$ on age profiles of Sa/Sb galaxies from \cite{Sanchez2020}, finding $\tausfh=15\Gyr$ at the solar annulus.  In their models the value of $\taustar$ at the solar radius is slightly over $3\Gyr$ at the present day, while a direct estimate of $\Sigma_{\rm SFR}/(\Sigma_{\rm HI}+\Sigma_{{\rm H}_2})$ using values from \cite{Elia2022}, \cite{Kalberla2009}, and \cite{Miville2017} gives $3.6\Gyr$.  The star formation efficiency may have been higher (shorter $\taustar$) in the past, when the galaxy was more gas rich, closer to the $2\Gyr$ timescale typical for molecular gas \citep{Leroy2008,Sun2023}.  
 These considerations suggest $0.13 < \taustar/\tausfh < 0.4$ as a plausible range.  

For $r=0.4$, $\taustar/\tausfh=0.2$, and $\yxcc/\Zxsun=0.973$ from \Eqref{yxccscale}, we find $\eta_\odot = 0.57$.  Lowering $\xfecc$ to 0.3 (implying $\yxcc/\Zxsun=0.689$) gives $\eta_\odot=0.29$, while raising $\xfecc$ to 0.6 ($\yxcc/\Zxsun=1.37$) gives $\eta_\odot=0.97$.  For our fiducial parameter choices, therefore, mild outflows with $\eta \approx 0.6$ are required to reach a solar ISM at last times.  With a low value of $\xfecc$ and uncertainties in $\ybarfecc$, $\Fexp$, and $\taustar/\tausfh$, a solution with no outflows ($\eta_\odot=0$) is possible.  With parameters pushed in the other direction --- $\xfecc=0.6$, $\ybarfecc=0.065$, $\Fexp=0.85$, $\taustar/\tausfh=0.4$ --- \Eqref{etasun}\ implies $\etasun=1.55$.  This is still lower than the value $\eta \approx 2.5$ adopted in the models of \cite{Andrews2017} and WAF because their theoretically motivated oxygen yield $\yocc = 0.015$ corresponds to $\yocc/\Zosun\approx 2$, twice the value suggested by the RMN23 determination of $\ybarfecc$.

For the $\taustar$ and $\tausfh$ values advocated by \cite{Johnson2021}, the timescale to reach equilibrium is quite short, $\taubar \sim 3\Gyr$, so the departures from equilibrium are expected to be small.  However, with a longer value of $\taustar$ these departures can become more significant.  In experimentation with the time-dependent solution (\Eqref{Zevol}) we find that the value of $\eta$ required to reach $Z_{\rm X}(t=13\Gyr) = \Zxsun$ is usually close to the value implied by \Eqref{etasun}\ even accounting for these departures, but caution is required if the denominators of Equations~(\ref{eqn:Zeq}) or~(\ref{eqn:taubar}) approach zero.  At fixed $\taustar$, a more sharply declining star formation history (shorter $\tausfh$) increases $Z_{\rm X}(t)$, so the required value of $\eta$ is higher even though the abundance is further below equilibrium. 

\subsection{Implications for the SNIa rate}
\label{sec:snia}

Once SNIa become an important enrichment channel, the value of $\afe$ in the ISM and newly forming stars falls below $\afecc$ because Fe now includes the additional SNIa contribution.  WAF approximated the delay time distribution (DTD) of SNIa as an exponential, $r_{\rm Ia}(t) \propto e^{-(t-t_d)/\tauIa}$, where $\tauIa \approx 1.5\Gyr$ and $t_d \approx 0.1\Gyr$ is the minimum delay time required to produce a SNIa.  The mass of iron produced by SNIa per unit mass of star formation is
\begin{equation}
\yfeIa = \RIa \ybarfeIa~,
\label{eqn:yfeIa}
\end{equation}
where $\RIa=N_{\rm Ia}/M_*$ is the time-integrated SNIa rate and $\ybarfeIa$ is the mean iron yield per Type Ia supernova.  We take $\ybarfeIa=0.7\Msun$, a typical value inferred from the mass of radioactive $^{56}$Ni in the analysis of \cite{Howell2009}, though estimated $^{56}$Ni masses span a wide range from supernova to supernova (e.g., \citealt{Childress2015}).  Like CCSN enrichment, SNIa enrichment also approaches equilibrium at late times, and one can use the ratio of equilibrium abundances (WAF Equations 29, 30) to show that
\begin{equation}
\afeeq = \afecc - \log\left(1+\mu \frac{\RIa\ybarfeIa}{\Rcc\ybarfecc}\right)~,
\label{eqn:afeeq}
\end{equation}
where 
\begin{equation}
\mu = \frac{\tausfh}{\tausfh-\tauIa} e^{t_d/\tausfh}
\label{eqn:mudef}
\end{equation}
is a constant that approaches unity for $\tausfh \gg \tauIa$.  \Eqref{afeeq}\ captures the expectation that the decrease of $\afe$ at late times depends on the rate of SNIa enrichment relative to CCSN enrichment, and the factor $\mu$ accounts for the fact that the SNIa rate is tied to the past star formation rate while CCSN are tied to the current star formation rate.  

We can rearrange \Eqref{afeeq}\ to solve for the SNIa rate,
\begin{equation}
\RIa = \frac{\Rcc}{\mu} \frac{\ybarfecc}{\ybarfeIa} \left(10^{\Delta\afe} - 1\right) ~,
\label{eqn:RIa}
\end{equation}
where $\Delta\afe = \afecc-\afeeq$ is the drop in $\afe$ between the CCSN plateau and the late-time equilibrium.
Taking $\mu=1.1$, $\ybarfecc/\ybarfeIa = 0.058/0.7 = 0.083$, $\Rcc = 0.008 M_\odot^{-1}$, $\afeeq=0$, and $\afecc=0.45$ gives $\RIa= 1.1\times 10^{-3} M_\odot^{-1}$.  The corresponding population-averaged yield is
\begin{multline}
    \yfeIa = \left(7.7 \times 10^{-4}\right) \times \left({\Fexp \over 0.75}\right)\left({\Fimf \over 1.0}\right)\left({\ybarfecc \over 0.058\Msun}\right) \\ \times \left({1.1 \over \mu}\right)
    \left({10^{\Delta\afe}-1 \over 1.82}\right)~,
    \label{eqn:yfeIascale}
\end{multline}
where we have restored the parameter dependencies to facilitate scaling to other choices.  Note that $\ybarfeIa$ cancels out of this abundance-based inference of $\yfeIa$, though the inferred $\RIa$ itself scales as $(0.7\Msun/\ybarfeIa)$.

\cite{Maoz2017} fit the observed cosmic star formation history and SNIa history to infer a $t^{-1.1}$ DTD with a Hubble-time integrated normalization $\RIa = (1.3 \pm 0.1) \times 10^{-3} M_\odot^{-1}$ for a Kroupa IMF.  Compared to an exponential DTD, a $t^{-1.1}$ DTD leads to slightly lower $\afe$ at late times, by roughly 0.05 dex (see Figure~11 of WAF).  However, the $\afe$ of recently formed stars in the solar neighborhood may also be slightly sub-solar, compensating this difference.  We conclude that, within the uncertainties, the value of $\RIa$ implied by RMN23's value of $\ybarfecc$ is consistent with the value found by \cite{Maoz2017}, assuming $\afecc=0.45$.  For $\afecc-\afeeq = 0.3$ or 0.6, the value of $\RIa$ implied by \Eqref{RIa}\ is lower by a factor of 1.8 or higher by a factor of 1.6, respectively, so the uncertainty in the true level of the $\afe$ plateau remains a substantial uncertainty in this inference of $\RIa$.  

\subsection{Implications for black hole formation}
\label{sec:s16}

For a given IMF, the population-averaged CCSN yield is sensitive to which massive stars actually explode.  We can therefore use the empirically inferred mean element yields to constrain the degree of black hole formation.  Here we examine the solar metallicity models of S16 as extended by G21.  S16 use the KEPLER code \citep{weaver1978} to compute the evolution of a dense grid of massive stars up to core collapse, then apply different models of neutrino-driven central engines to compute the subsequent explosion or implosion.  G21 extend this suite by forcing explosions at all masses, with explosion energies and the boundary between ejected and fallback material calibrated by the neutrino-driven engine results.  With this grid of mass-dependent yields, one can impose a black hole formation landscape {\it a posteriori} and compute the IMF-averaged element yields from the CCSN and the pre-supernova stellar winds.

\begin{figure*}[!htb]
    \centering
    \includegraphics[width=\textwidth]{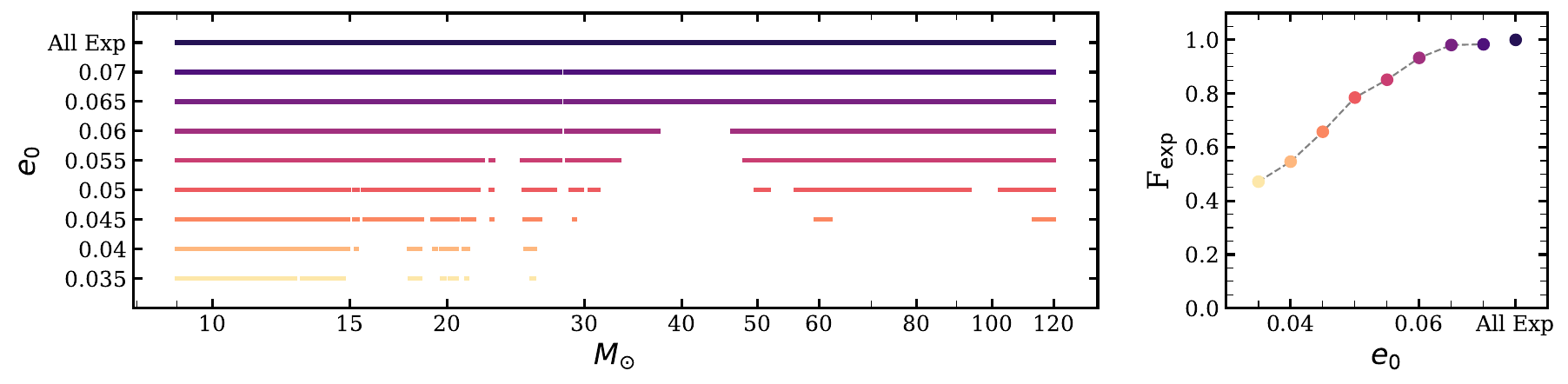}
    \caption{Left: Continuous explosion landscapes for a range of $e_0$ values ($0.035-0.07$ and All Explode) as a function of progenitor ZAMS mass. Horizontal lines indicate the masses where successful explosions occur. Lines are colored by $e_0$, with low values of $e_0$ in yellow and high values of $e_0$ in dark purple. While the $e_0=0.065$ and $e_0=0.07$ cases appear to fully explode, there are small zones of black hole formation.  Right: $\Fexp$ for the $e_0$ landscapes shown in the left panel.}
    \label{fig:eofm}
\end{figure*}

For the same models as S16, \cite{Ertl2016} find that exploding and non-exploding massive star progenitors can be well separated by a critical curve in the space of $\mu_4$, $M_4$ parameters linked to the mass infall rate and neutrino luminosity at core collapse.  Based on these results, G21 define an explodability function
\begin{equation}
    E(M,e_0) = \Theta\left(0.28 M_4 \mu_4 - \mu_4 + e_0\right) ~,
    \label{eqn:explodability}
\end{equation}
where $\Theta$ is the Heaviside step function and progenitors with $E(M,e_0)=1$ explode and $E(M,e_0)=0$ implode.
The quantity $e_0$ encodes the power of the central engine. For $e_0 \geq 0.065$ nearly all massive stars explode, and for $e_0 \leq 0.03$ only stars with $M \leq 14\Msun$ and a handful of larger masses explode.  At intermediate values of $e_0$ the $E(M,e_0)$ function traces a complex landscape of black hole formation because the structure of the pre-supernova progenitor is a sensitive, non-monotonic function of the initial mass $M$ (\citealt{Pejcha2015,Ertl2016}; S16; \citealt{Ebinger2019}).  Figure~\ref{fig:eofm}, similar to Figure~5 of G21, illustrates the dependence of the explosion landscape and the $\Fexp$ value on $e_0$.  The ``Z9.6+W18'' neutrino-driven model that S16 adopt as representative corresponds to $e_0=0.043$.  

For each explosion landscape we calculate the IMF-averaged net yield, 
\begin{equation}
  \yxcc = \dfrac{
     \int_{\Msn}^{\Mmax} \left[E(M,e_0)\mxexp + \mxwind - b_{\rm X}\Delta M \right]{\frac{dN}{dN}dM}
  }
    {\int_{\Mmin} ^{\Mmax} M \frac{dN}{dM} dM} ~.
\end{equation}
Here $\Delta M = M-\Mrem$ is the difference between the star's zero-age main sequence (ZAMS) mass and that of its neutron star or black hole remnant, and $\mxexp$ and $\mxwind$ are the mass of element X ejected in the explosion and the pre-supernova wind, respectively.  We subtract the ejected mass $b_X \Delta M$ that was present in the star at birth to obtain the net yield of newly produced material.  We again adopt a Kroupa IMF with $\Mmin=0.08M_\odot$ and $\Mmax=120 M_\odot$, and we assume a minimum CCSN progenitor mass $\Msn=8M_\odot$.  We only include the wind component for O, as massive star winds do not carry newly produced Mg, Si, or Fe.

We compute $M_{\rm rem}$ from the information tabulated by S16/G21 by subtracting the sum of {\it all} element yields, including hydrogen and helium, from the ZAMS mass:
\begin{equation}
    M_{\rm rem} = M - \Sigma \mxwind - E(M,e_0)(\Sigma \mxexp)~.
    \label{eqn:Mrem}
\end{equation}
For a non-exploding star, with $E(M,e_0)=0$, the remnant black hole contains all mass that was not ejected in winds.  For exploding models our calculated values of $M_{\rm rem}$ agree with the published values from S16.

\begin{figure}
    \centering
    \includegraphics[width=\columnwidth]{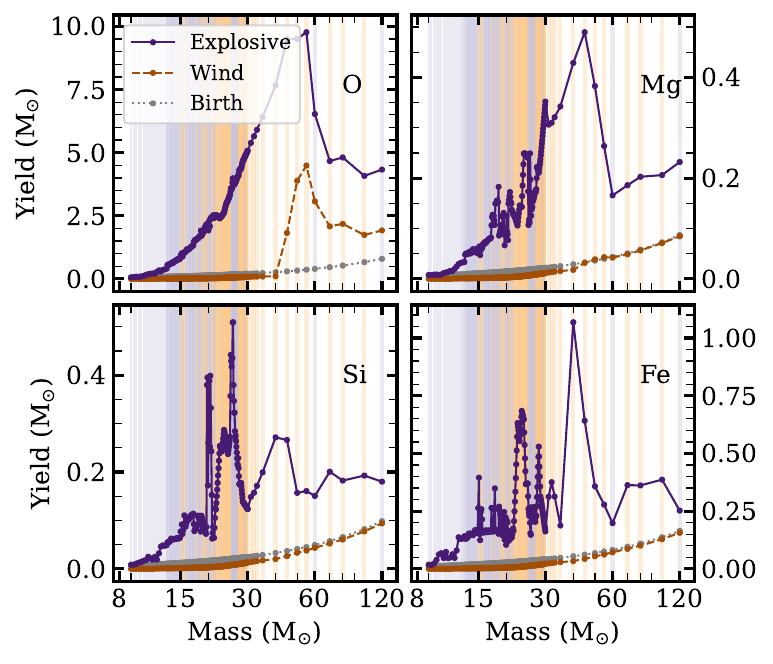}
    \caption{Explosive (dark purple) and wind (dark orange) yields from S16 as a function of progenitor mass for elements O (top left) Mg (top right), Si (bottom left), and Fe (bottom right). We plot the progenitor birth abundances in grey. The background colored lines indicate stellar models that explode (light purple) and collapse (light orange) for a landscape with $e_0=0.046$.  For this $e_0$, which gives the best agreement with the RMN23 value of $\ybarfecc=0.058M_\odot$, only the masses marked by purple lines would explode and release newly synthesized Fe.}
    \label{fig:yields}
\end{figure}

Figure~\ref{fig:yields} plots the S16/G21 yields of O, Mg, Si, and Fe as a function of progenitor mass, similar to Figure 2 of G21.  G21 force explosions for all masses on this fine grid.  Vertical orange lines indicate masses that would implode (and thus produce no explosive yield) for a landscape with explodability threshold $e_0=0.046$.  For O, the predicted yield rises steadily and steeply up to $M=60M_\odot$.  The S16 models adopt an aggressive mass loss prescription that strips the envelopes of stars with $M>40M_\odot$.  Above this mass the wind yields of O are significant, but high mass progenitors have reduced explosive yields because of their early mass loss.  The sum of wind and explosive yields is roughly constant at about $7M_\odot$ for $M>70M_\odot$.  Because of IMF-weighting, stars with $M>40M_\odot$ have only moderate impact ($\sim 20\%$) on the IMF-averaged O yield, so although the uncertainties in mass loss are substantial, they do not drastically affect the predicted $\yocc$.  

The behavior for Mg is similar to O, but at intermediate masses there are spikes in yield over narrow progenitor mass ranges that produce denser pre-supernova cores, which create additional Mg during the explosion.  The Si yield is dominated by explosive nucleosynthesis rather than hydrostatic (pre-explosion) nucleosynthesis, so this spikiness is more pronounced and the overall mass trend is weaker.  The sharp, non-monotonic variation is even stronger for the Fe yield.  Progenitor models near the threshold of explodability tend to produce high Fe yields if they {\it do} explode because they are characterized by dense cores that can approach nuclear statistical equilibrium once explosive nucleosynthesis takes place.  The properties of pre-supernova cores can change sharply with small changes in ZAMS progenitor mass because of shifts in the spatial position of nuclear fusion zones (S16; \citealt{Sukhbold2018}).  

\begin{figure}[!htb]
    \centering
    \includegraphics[width=\columnwidth]{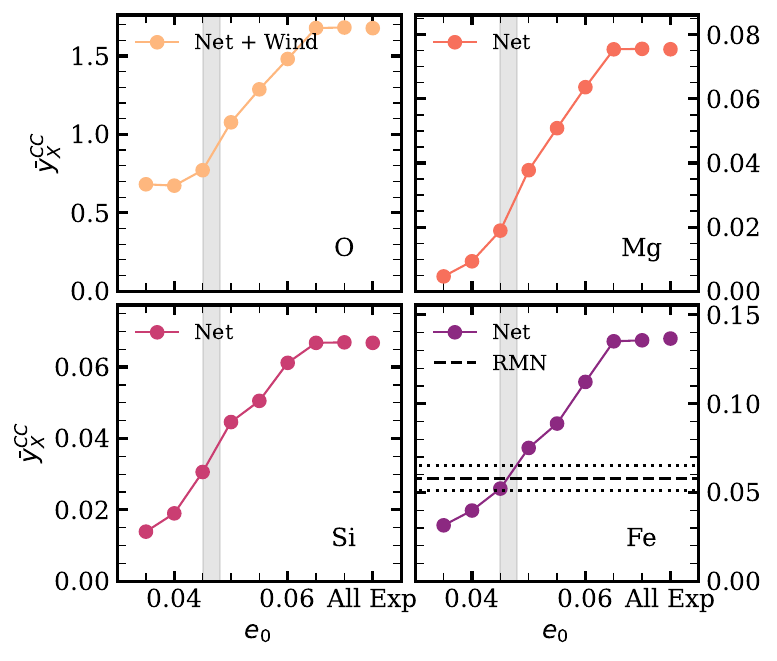}
    \caption{Average yield per supernova $\ybarxcc$, in $\Msun$, as a function of explodability threshold for $e_0 = 0.035-0.07$ + All Explode for O (top left), Mg (top right), Si (bottom left), and Fe (bottom right). In the bottom right panel for Fe we also plot the RMN23 value of $\ybarfecc=0.058\pm 0.007\Msun$, and in each panel we shade the region between $e_0=0.045-0.048$, the $e_0$ values in agreement with the RMN23 Fe yield. }
    \label{fig:avg_yields}
\end{figure}

Table~\ref{tbl:yield} lists the explosion fraction $\Fexp$ and the population-averaged net yields $\yfecc$, $\yocc$, $\ymgcc$, and $\ysicc$ computed from the S16/G21 models as a function of the explodability threshold $e_0$, for a Kroupa IMF.  These yields are dimensionless, representing solar masses of element production per solar mass of star formation.  We also list the mean yield per supernova, which from Equations~(\ref{eqn:ncc}) and~(\ref{eqn:yxcc}) is
\begin{equation}
    \ybarxcc = {\yxcc \over \Rcc} = 91.7 \Fexp^{-1} y_{\rm X,exp}^{\rm cc} \,\Msun~,
    \label{eqn:ybar}
\end{equation}
where we assume $\Fimf=1$. For Mg, Si, and Fe, $y_{\rm X,exp}^{\rm cc} = \yxcc$, but for O we omit the wind yield from non-exploding stars so that $\ybarocc$ represents the average yield of the actual CCSN.  Figure~\ref{fig:avg_yields} plots these average yields as a function of $e_0$.  
As illustrated in the lower right panel, the RMN23 value of $\ybarfecc = 0.058 \pm 0.007\Msun$  leads to a fairly tight constraint $0.045 < e_0 < 0.048$.  The best-fit value $e_0=0.046$ is above the value corresponding to the Z9.6+W18 model of S16 and very close to that of their Z9.6+N20 model.  The orange lines in Figure~\ref{fig:yields} mark the progenitor masses that would implode to black holes without releasing any newly synthesized Fe for this value of $e_0$.

\begin{deluxetable*}{cccccccccc}
    \tablecaption{CCSN Yields From the S16/G21 Models \label{tbl:yield}}
    \tablehead{
    \colhead{$e_0$} & \colhead{$\Fexp$} & \colhead{$\yocc$} & \colhead{$\ymgcc$} &  \colhead{$\ysicc$} & \colhead{$\yfecc$} &
    \colhead{$\ybarocc$} & \colhead{$\ybarmgcc$} & \colhead{$\ybarsicc$} & \colhead{$\ybarfecc$}}
    \startdata
    0.035 &	0.472 &	3.51e-03 &	2.38e-05 &	7.15e-05 &	1.62e-04 &	6.82e-01 &	4.62e-03 &	1.39e-02 &	3.15e-02 \\
    0.040 &	0.546 &	4.01e-03 &	5.56e-05 &	1.13e-04 &	2.37e-04 &	6.73e-01 &	9.34e-03 &	1.90e-02 &	3.98e-02 \\
    0.045 &	0.658 &	5.54e-03 &	1.36e-04 &	2.20e-04 &	3.74e-04 &	7.72e-01 &	1.89e-02 &	3.06e-02 &	5.22e-02 \\
    0.050 &	0.785 &	9.23e-03 &	3.23e-04 &	3.82e-04 &	6.43e-04 &	1.08e+00 &	3.78e-02 &	4.46e-02 &	7.51e-02 \\
    0.055 &	0.851 &	1.20e-02 &	4.73e-04 &	4.69e-04 &	8.25e-04 &	1.29e+00 &	5.09e-02 &	5.05e-02 &	8.88e-02 \\
    0.060 &	0.933 &	1.51e-02 &	6.48e-04 &	6.23e-04 &	1.14e-03 &	1.48e+00 &	6.37e-02 &	6.12e-02 &	1.12e-01 \\
    0.065 &	0.981 &	1.80e-02 &	8.08e-04 &	7.15e-04 &	1.45e-03 &	1.68e+00 &	7.55e-02 &	6.68e-02 &	1.35e-01 \\
    0.070 &	0.983 &	1.80e-02 &	8.11e-04 &	7.18e-04 &	1.46e-03 &	1.68e+00 &	7.56e-02 &	6.69e-02 &	1.36e-01 \\
    All Exp & 1.000 & 1.83e-02 & 8.24e-04 &	7.29e-04 &	1.49e-03 &	1.68e+00 &	7.55e-02 &	6.68e-02 &	1.37e-01 \\
    \enddata
    \tablecomments{For the $e_0=0.035-0.07$ and All Explode CCSN landscapes, columns list $\Fexp$, net yield $\yxcc$ in $\Msun$ per $\Msun$ formed, and the average net explosive yield per supernova, $\ybarxcc$.  For O, the $\yxcc$ calculation includes wind contributions but the $\ybarxcc$ calculation does not.  Wind contributions to the net yields are negligible for Mg, Si, and Fe.}
\end{deluxetable*}

At the best value of $e_0$, the predicted O, Mg, and Si yields are $\ybarocc=0.83\Msun$, $\ybarmgcc=0.023\Msun$, and $\ybarsicc=0.033\Msun$, respectively.    These can be compared to our empirically inferred values (\Eqref{ybarcc}) of
$0.87\Msun$, $0.080\Msun$, $0.082\Msun$.  For the $e_0$ value implied by Fe, the predicted O yield is in good agreement with our empirical inference, but the predicted Mg and Si yields are low by factors of 3.5 and 2.5, respectively. 

This inconsistency among O, Mg, and Si is already evident in the relative yields of the models, as shown by G21 (see their Figure~9). For any choice of explosion landscape, these models overpredict the solar O/Mg ratio by a factor of 2.5-4, much larger than the observational uncertainty in this ratio. For a landscape like Z9.6+N20 that gives agreement with the observed $\ybarfecc$, the models also overpredict the solar Si/Mg ratio (after accounting for the SNIa contribution to Si) by a factor $\sim 2$.  Even if we adopted a high value of $e_0$ for which essentially all massive stars explode, the Mg and Si yields would remain (slightly) below our empirically inferred values.  We caution, however, that our empirical values rely on our uncertain choice of $\afecc = 0.45$.  If we lower $\afecc$ to 0.3, then the inferred $\ybarmgcc$ and $\ybarsicc$ drop to $0.056\Msun$ and $0.058\Msun$, respectively, which the S16/G21 models would produce for $e_0 \approx 0.057$.  However, with this $e_0$ the models overproduce $\ybarfecc$ and $\ybarocc$ by factors of $\sim 1.5-2$.  

It is encouraging that the S16 models with a physically plausible choice of neutrino-driven engine can reproduce both the RMN23 Fe yield and our inferred O yield, but the failure to reproduce Mg and Si tempers our assessment of this success.
In comparing the S16/G21 models to our empirically inferred yields, we implicitly assume that these models represent all CCSN types including the Ib and Ic supernovae that RMN23 include in their $\ybarfecc$ determination.  It is possible that Ib and Ic supernovae have different progenitor structure because of hydrogen envelope stripping and therefore require separate treatment in yield calculations \citep{Sukhbold2020,Laplace2021}.

\subsection{Implications for ISM deuterium}
\label{sec:deuterium}

Deuterium is the one isotope whose nucleosynthetic yield is securely and precisely predicted by theory: stars destroy all the D they are born with because in their fully convective proto-stellar phase they cycle their birth D through layers hot enough to fuse it into $^4$He \citep{Bodenheimer1966,Mazzitelli1980}.  All D present in the ISM must have originated in the big bang, and it resides in fluid elements that were never processed through stars.  \cite{Weinberg2017deut} shows that, in a variety of GCE models, the evolution of the ISM D/H ratio $\Xd$ is accurately approximated by
\begin{equation}
    \Xd = {\Xdp \over 1+r Z_{\rm X}/\yxcc}~,
    \label{eqn:xd}
\end{equation}
where $\Xdp$ is the primordial D/H and $Z_{\rm X}$ is the mass fraction of a pure CCSN element with metallicity independent yield $\yxcc$.  \cite{vandevoort2020} show that \Eqref{xd}\ also accurately describes the results of hydrodynamic cosmological simulations.  While the form of \Eqref{xd}\ is motivated by analytic solutions at equilibrium, the approximate relation follows from basic considerations.  A stellar population of mass $M_*$ produces a mass $\yxcc M_*$ of element X and a mass $rM_*$ of gas, so its ejecta have a mean $Z_{\rm X} = \yxcc/r$.  The ratio $\Xd/\Xdp$ tells what fraction of the ISM is primordial (never processed through stars), and thus the factor by which $Z_{\rm X}$ is diluted relative to the ratio in stellar ejecta.

Taking $\ybarxcc/\Zxsun=0.973$ and $r=0.4$, we predict $\Xd/\Xdp=0.71$ for solar metallicity.  With primordial D/H of 26 ppm \citep{Cyburt2016} this implies ISM D/H of about 18.5 ppm.  \cite{Linsky2006} measure D/H in absorption along many lines of sight through the ISM finding values that span a factor of three.  They attribute this variation to depletion of D onto small dust grains along some lines of sight, and based on the highest D/H (least depleted) sightlines they advocate an ISM abundance $23.1 \pm 2.4$ ppm.  This is higher than the value predicted at solar metallicity for our fiducial yield scale, but consistent at $2\sigma$.  Reducing the population-averaged $\alpha$-element yield significantly below $Z_{\alpha,\odot}$ would make the high D/H values from \cite{Linsky2006} difficult to reproduce within conventional GCE models.  For example, if we adopt the ratio $y_{\alpha}^{\rm cc}/Z_{\alpha,\odot} = 0.689$ implied by \Eqref{yxccscale}\ for $\afecc=0.3$, then the predicted ISM D/H falls to 16.4 ppm, a $2.8\sigma$ conflict with Linsky et al.'s value.

\section{Discussion}
\label{sec:discussion}

\subsection{Overview}
\label{sec:overview}

\begin{figure*}
    \centering
    \includegraphics[width=\textwidth]{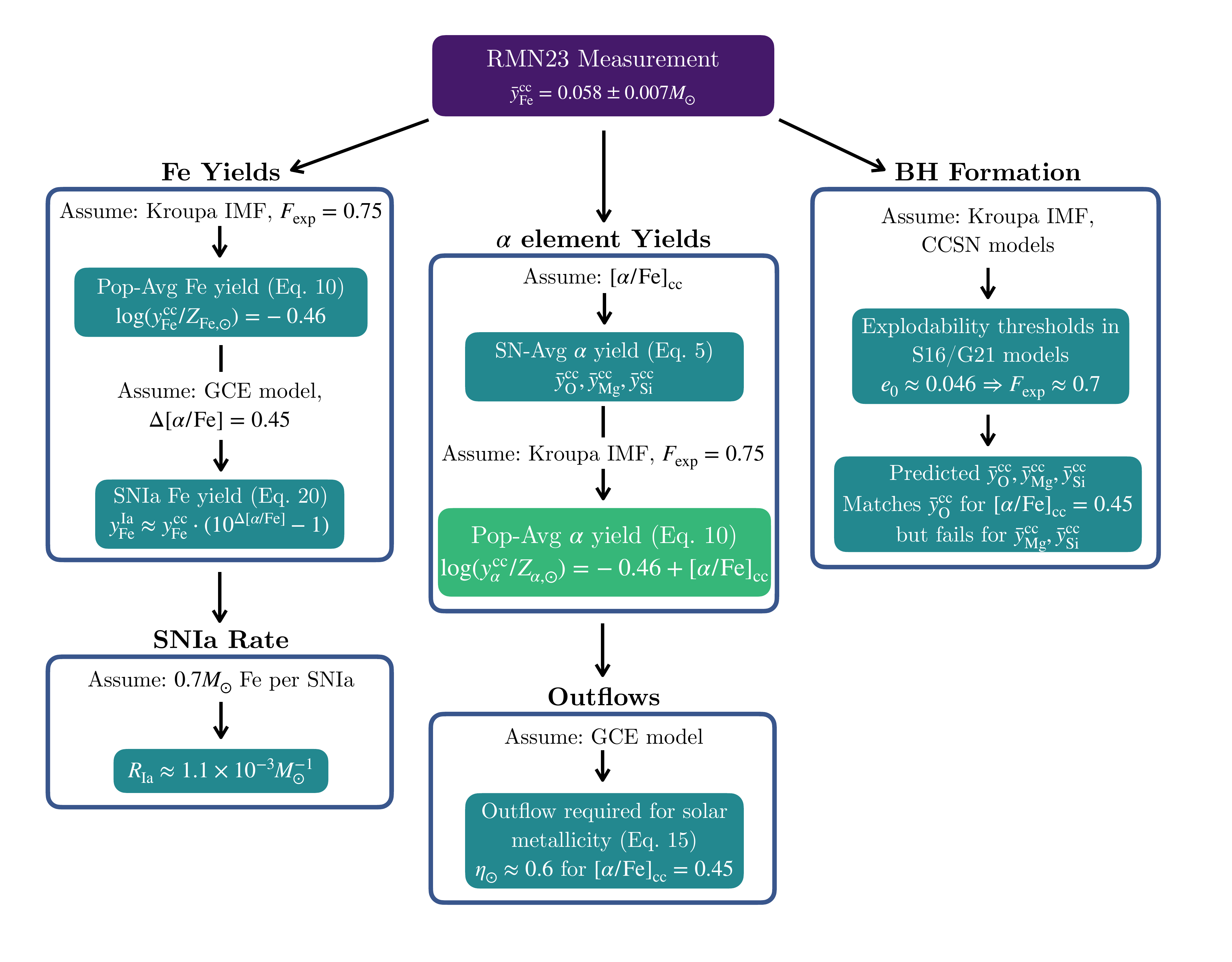}
    \caption{Overview of our results, indicating where parameter values and GCE or CCSN models enter the chain of inference.  The central column traces our core results on the scale of $\alpha$-element yields and its implication for outflows.  The left column traces conclusions about Fe yields from CCSN and SNIa.  The right column traces implications for the S16/G21 models of CCSN and black hole formation. }
    \label{fig:summary}
\end{figure*}

Figure~\ref{fig:summary} presents an overview of our findings, with an emphasis on where different assumptions enter the logical flow.  The center column traces the route to our central result, the population-averaged yield of $\alpha$-elements (\Eqref{yxccscale}).  The assumed value of $\afecc$ is sufficient to translate the RMN23 measurement of the mean Fe yield per CCSN to the corresponding mean yields of $\alpha$ elements (\Eqref{ybarcc}).  Converting to the population-averaged yields (solar masses produced per solar mass of star formation) requires an assumed IMF and explosion fraction.  For our fiducial scalings, \Eqref{yxccscale}\ reduces to
\begin{equation}
    \log{y_\alpha^{\rm cc} \over Z_{\alpha,\odot}} = -0.46 + \afecc ~.
     \label{eqn:logyalpha}
\end{equation}
Thus, for $\ofecc=\mgfecc=0.45$, the implied yields of O and Mg are almost exactly equal to their solar mass fractions.  Conclusions about outflows rely on a GCE model in combination with these yields.  For straightforward assumptions, a mass-loading factor $\eta \approx 0.6$ is required to yield a solar metallicity ISM at late times (\Eqref{etasun}).

The left column traces our conclusions about Fe yields.  Deriving a population-averaged CCSN Fe yield from the RMN23 measurement requires values of $\Fimf$ and $\Fexp$, but it does not depend on $\afecc$ because $[{\rm Fe}/{\rm Fe}]_{\rm cc}=0$ by definition.  Inferring the population-averaged SNIa Fe yield from the corresponding CCSN yield requires a GCE model and a value for the gap $\Delta\afe$ between the CCSN plateau and the late-time equilibrium (\Eqref{yfeIascale}).  For $\afecc=0.45$ and reasonable choices of star formation history and SNIa DTD, a model with $\yfeIa \approx 1.65\yfecc$ evolves to $\afe \approx 0$ at late times.  With an empirically and theoretically motivated choice of the mean Fe yield per SNIa, $\ybarfeIa=0.7\Msun$, one can infer the Hubble-time integrated SNIa rate $\RIa \approx 1.1\times 10^{-3} \Msun^{-1}$.

The right column traces the implications of the RMN23 measurement for the S16/G21 models of CCSN and black hole formation.  The measured $\ybarfecc$ and the assumption of a Kroupa IMF leads directly to a constraint on the explodability threshold in these models, $e_0 \approx 0.046$, with an explosion landscape similar to the 3rd-from-bottom line in Figure~\ref{fig:eofm}.  In this model, the fraction of $M>8\Msun$ stars that explode as CCSN is $\Fexp \approx 0.7$.  With $e_0$ fixed by matching the observed $\ybarfecc$, the model predicts the SN-averaged yields of O, Mg, and Si (Table~\ref{tbl:yield}).  For O, the predicted yield of $\ybarocc=0.83\Msun$ agrees well with the empirically inferred value for $\afecc=0.45$, but Mg and Si are underpredicted by a factor $\sim 3$.  Reproducing the inferred Mg and Si yields would require nearly all $M>8M_\odot$ stars to explode, but the model would then drastically overpredict O.  This conflict among O, Mg, and Si is likely rooted within the S16/G21 models themselves rather than choices we have made here, perhaps reflecting inaccuracies in the adopted nuclear reaction rates (see G21 for further discussion).  

\subsection{Uncertainties}
\label{sec:uncertainties}

The single largest uncertainty in deriving $\yxcc$ from the RMN23 measurement of $\ybarfecc$ is the choice of $\xfecc$, the ratio of $\alpha$-element production to Fe production by massive stars.  We have chosen $\ofecc=\mgfecc=0.45$ as our fiducial value based on the plateau in $\afe$ vs. $\feh$ observed for stars with $\feh \leq -0.8$.  However, the observed level of the plateau varies from study to study and in some cases depends on the choice of $\alpha$-element \citep{Kobayashi2020}.  The RMN23 sample is likely dominated by CCSN progenitors near solar metallicity because galaxies with $L \sim L_*$ and $Z \sim Z_\odot$ contribute most to the global star formation rate.  Taking the low-metallicity plateau to represent $\afecc$ for the RMN23 supernovae implicitly assumes that this ratio does not change between $\feh \approx -1$ and $\feh \approx 0$.
This assumption is reasonable because the predicted yields of these elements do not depend strongly on metallicity (see \citealt{Andrews2017}, Figure 20).  However, these predictions could be incorrect if black hole formation or the stellar IMF change systematically with metallicity in a way that favors Fe or $\alpha$-element production.

\begin{figure}
    \centering
    \includegraphics[width=\columnwidth]{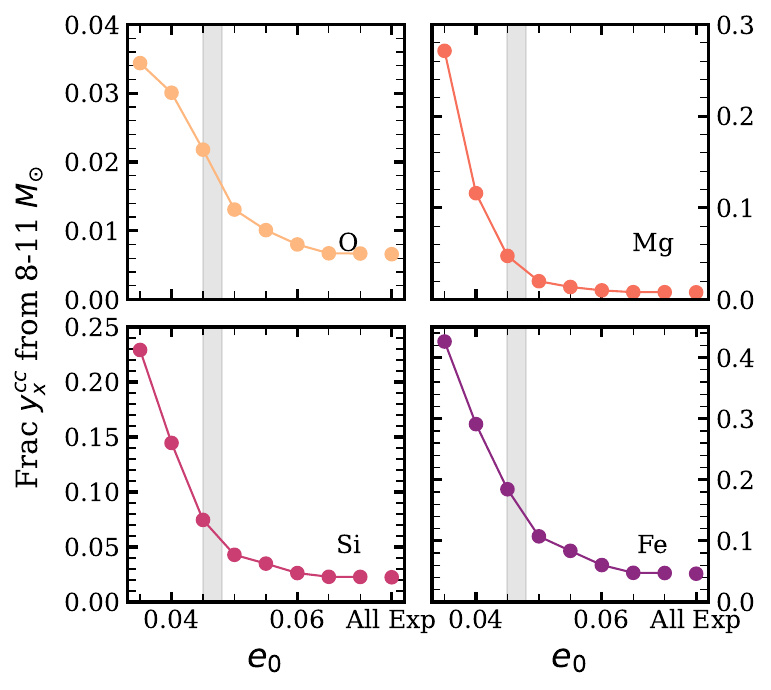}
    \caption{Fraction of IMF-averaged yield contributed by stars with $M=8-11\Msun$, as a function of explodability threshold, for elements O (top left), Mg (top right), Si (bottom left), and Fe (bottom right). We shade the region between $e_0 = 0.045 - 0.048$, the $e_0$ values in agreement with the RMN23 Fe yield.  Note that the $y$-axis scale is different for each panel.}
    \label{fig:yield8-11}
\end{figure}

RMN23 discuss a number of sources of uncertainty in their $\ybarfecc$ measurement, and we assume that these are adequately reflected in their $0.007\Msun$ error bar.  One specific concern is that the RMN23 sample might undercount low luminosity supernovae from low mass progenitors, which have lower than average $^{56}$Ni production and thus lower than average Fe yields.  However, these faint supernovae also have lower than average $\alpha$-element yields, so the ratio $\ybarxcc/\ybarfecc$ may not change much by omitting them. To illustrate this point, Figure~\ref{fig:yield8-11} plots the fraction of O, Mg, Si, and Fe produced by CCSN with progenitor mass $M=8-11\Msun$, as a function of the explodability threshold $e_0$, using the S16/G21 yields.  For the $e_0=0.046$ favored by our analysis in \S\ref{sec:s16}, these low mass progenitors produce only 15\% of the Fe, 3-5\% of the Mg and Si, and 1.5\% of the O.  These moderate fractions imply that underrepresentation of these faint CCSN in the RMN23 sample would have limited impact on our yield conclusions.  The impact on the comparison to the S16/G21 models is more complex, because if the $8-11\Msun$ mass range is omitted then $e_0$ must be lowered (to $e_0 \approx 0.4$) to reproduce $\ybarfecc=0.058\Msun$, which changes the predicted $\Fexp$ and yield ratios.  

\Eqref{yxccscale} expresses the dependence of our inferred $\alpha$-element yield scale on the quantities $\Fexp$, $\Fimf$, and $\xfecc$, so the equation itself is model-independent.  To assign an approximate error bar to our $\yxcc$ values in \Eqref{yxccvals}, we take $\afecc=0.45 \pm 0.15$ to represent the ``$\pm 2\sigma$'' range of plausible values for this quantity, making the ``$1\sigma$'' error bar 0.075 dex.  We take $\Fexp = 0.75 \pm 0.1$.  We then add the 0.075 dex uncertainty in $\afecc$ in quadrature with the 0.05 dex uncertainty ($\approx \log[1+0.1/0.75]$) in $\Fexp$ and 0.05 dex uncertainty ($= \log[1+0.007/0.058]$) of the RMN23 $\ybarfecc$ measurement to obtain
\begin{equation}
    \log y_{\alpha}^{\rm cc}/Z_{\alpha,\odot} = -0.01 \pm 0.1
    \label{eqn:yalpha}
\end{equation}
for the $\alpha$-elements O and Mg.  This error bar does not incorporate any uncertainty in the IMF, and in general one should use \Eqref{yxccscale}\ to account for specified changes in model assumptions.

\subsection{Outflows and the mass-metallicity relation}
\label{sec:disc_outflow}

For our fiducial choices of parameters, the inferred oxygen yield $\yocc \approx \Zosun \approx 0.007$ is substantially lower than the value $\yocc \approx 0.015$ assumed in many earlier papers by our group (e.g., \citealt{Andrews2017}; WAF; \citealt{Johnson2021}).  The higher $\yocc$, based on the Chieffi \& Limongi massive star models \citep{Chieffi2004,Chieffi2013,Limongi2006,Limongi2018}, a \cite{Kroupa2001} IMF, and minimal suppression of yield by black hole formation, is similar to that adopted in much of the literature on the galaxy mass-metallicity relation (e.g., \citealt{Finlator2008,Peeples2011,Dave2012,Zahid2012}).  With the lower yield inferred here, the efficiency of outflows required to reproduce observed galaxy metallicities will be lower, by roughly a factor of two in the $\eta \gg 1$ regime that applies at low halo masses.  Similarly, the average fraction of metals ejected by galaxies will be lower than the value of $\sim 75\%$ found by \cite{Peeples2014} because that was based on comparing the total metals produced by a galaxy's stellar population to the total metals remaining in the stars and ISM.  

\begin{figure}
    \centering
    \includegraphics[width=\columnwidth]{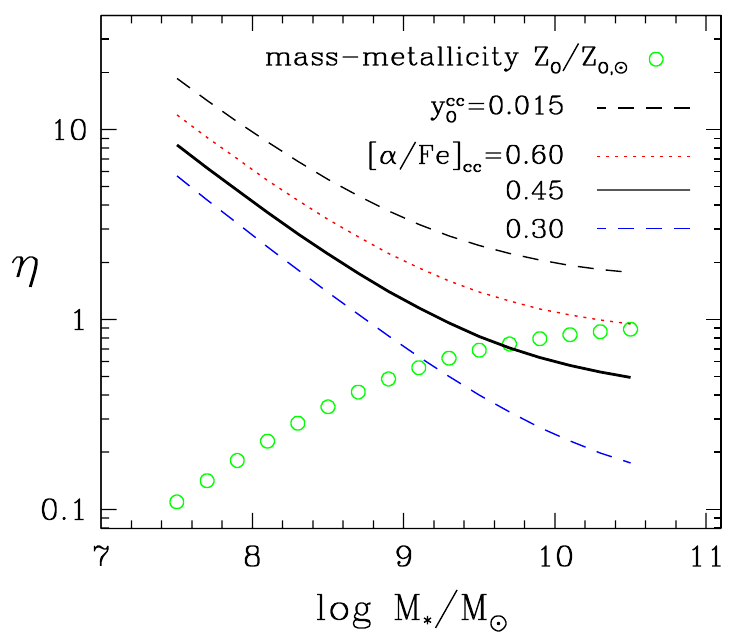}
    \caption{Outflow mass-loading required to reproduce the observed mass-metallicity relation at equilibrium.  Green points show the observed gas-phase oxygen abundance as a function of stellar mass derived from stacked SDSS spectra \citep{Andrews2013}, converted to $Z_{\rm O}/\Zosun$ using the \cite{Magg2022} abundance scale.  Curves show the values of $\eta$ that produce this abundance for a CCSN O yield $\yocc=0.015$ (black dashed) or for our empirical yield scaling with three different values of $\afecc$ (as labeled).}
    \label{fig:mzr}
\end{figure}

Figure~\ref{fig:mzr} illustrates the impact of lower yields on inferred outflow mass-loading, for the simplified assumptions of equilibrium abundances and constant SFR.  Open circles show the \cite{Andrews2013} fit to the observed relation between ISM oxygen abundance and galaxy stellar mass, which they derive from auroral lines in stacked SDSS spectra.  Over the galaxy stellar mass range $M_*=10^{7.5}\Msun-10^{10.5}\Msun$ these abundances increase from $\approx 0.1\Zosun$ to $\approx \Zosun$.  The solid black curve shows the value of $\eta$ required to produce the observed oxygen abundance in equilibrium for our fiducial yield scale, $\yocc=0.973\Zosun$, and assuming a constant SFR (\Eqref{Zeq}\ with $r=0.4$ and $\taustar/\tausfh=0$).  Red dotted and blue dashed curves show the corresponding results for $\afecc=0.6$ or 0.3, respectively.  The black dashed curve shows the inferred $\eta$ using a yield $\yocc=0.015=2.1\Zosun$ characteristic of studies based on the Limongi and Chieffi massive star yields with minimal black hole formation.  

The modeling assumptions adopted in Figure~\ref{fig:mzr} are idealized, and the observational relation has significant uncertainties, but the main implications are robust: large outflow mass-loading is needed to explain the low ISM metallicities of low mass galaxies, but the required $\eta(M_*)$ is sensitive to the adopted yield scale.  For $M_* < 10^9M_\odot$ the fiducial curve in Figure~\ref{fig:mzr} is well described by
\begin{equation}
    \eta \approx  4\left({M_* \over 10^8 \Msun}\right)^{-0.6} \approx 
                4\left({M_{h,{\rm peak}} \over 6 \times 10^{10}\Msun} \right)^{-1.2}~, 
\label{eqn:etascale}
\end{equation}
where the second $\approx$ adopts the $z=0.1$ relation between stellar mass and peak halo mass from Figure~9 of \cite{Behroozi2019}.  The \cite{Andrews2013} gas-phase measurements cut off at $M_* \approx 10^{7.5}\Msun$.  Extrapolating to $M_*=10^6\Msun$, Equations~(\ref{eqn:etascale}) and~(\ref{eqn:Zeq}) predict $[\alpha/{\rm H}] \approx -1.8$, in reasonable agreement with the mean {\it stellar} metallicity found for galaxies in this mass range by \cite{Kirby2013}.  However, this measurement is [Fe/H] rather than [$\alpha$/H], and the relation $Z \propto M_*^{0.6}$ implied by \Eqref{etascale}\ at low masses is steeper than the stellar mass-metallicity scaling $Z_* \propto M_*^{0.3}$ found by \cite{Kirby2013}. Discussion of the outflows required to reproduce the full stellar metallicity distributions of individual dwarf galaxies, and the degeneracy between outflows and yields in this modeling, can be found in \cite{Johnson2022dwarf} and \cite{Sandford2022}.

Returning to the Milky Way regime, we find in \S\ref{sec:outflows} that $\eta \approx 0.6$ is required to produce a solar metallicity ISM at equilibrium, for an empirically plausible choice of $\taustar/\tausfh$.  Our GCE model assumes that ejected material has the same metallicity as the ISM, and if winds were metal-enhanced then the mass outflow would be lower while the metal outflow would be similar.  Even with $\yocc \approx \Zosun$ as found here, it is difficult to reproduce Milky Way disk abundances with no outflows, though it is certainly easier than it would be for $\yocc = 2-3\Zosun$.  We will examine this issue more thoroughly in future work that compares multi-zone GCE models to the Milky Way's observed gas and stellar abundance gradients, for a variety of assumptions about outflows and radial gas flows (J.W.\ Johnson et al., in preparation).  

\subsection{Milky Way chemical evolution}
\label{sec:MW}

\begin{figure}
    \centering
    \includegraphics[width=\columnwidth]{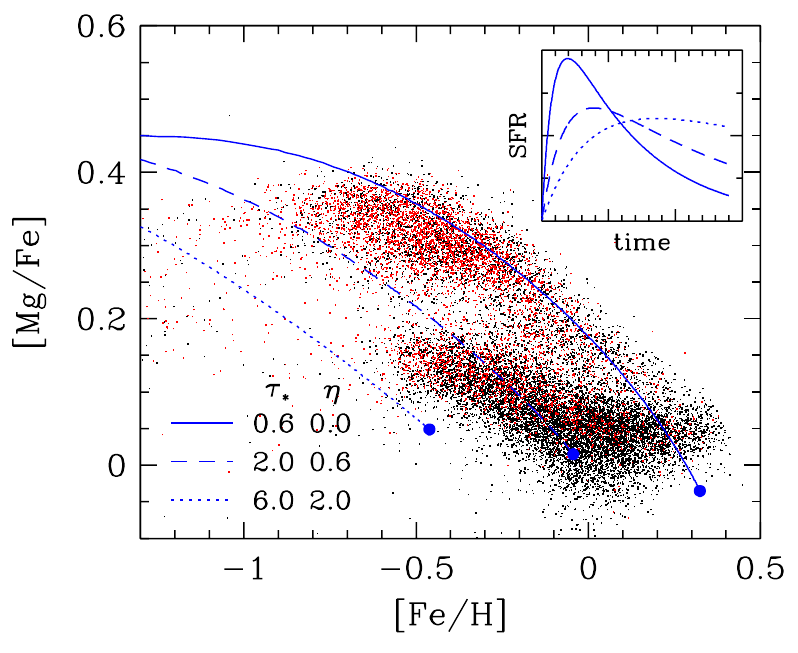}
    \caption{Evolutionary tracks of one-zone GCE models compared to APOGEE stellar abundance data from the solar annulus.  Points show $\mgfe$ vs. $\feh$ for APOGEE stars with $\Rgal=7-9\kpc$ and midplane distance $|Z|< 1\kpc$ (black, downsampled by a factor of four) or $|Z|=1-2\kpc$ (red).  Curves show models with the star formation histories shown in the inset and with SFE and outflow parameters $\taustar,\eta = 0.5, 0$ (solid), $2,0.6$ (dashed), or $6,2$ (dotted).
    }
    \label{fig:apogee2}
\end{figure}

In the $\afe-\feh$ plane, stars in the Milky Way thick disk and thin disk follow distinct ``high-$\alpha$'' and ``low-$\alpha$'' sequences \citep{Fuhrmann1998,Bensby2003}.  In the scenario proposed by \cite{Schoenrich2009}, the low-$\alpha$ sequence is not an evolutionary track itself but the superposition of endpoints of evolutionary tracks, with radial migration mixing stellar populations that formed at different Galactocentric radii.  As a very simple illustration of this picture, Figure~\ref{fig:apogee2} superposes three one-zone evolutionary tracks on the $\afe-\feh$ distribution of APOGEE stars in the solar annulus.  This Figure is similar to Figure~16 of WAF (and Figures 15 and 17 of \citealt{Nidever2014}), but with models adjusted to the lower yields of Equations~(\ref{eqn:yxccscale}) and~(\ref{eqn:yfeIascale}).  We also adopt the 2-parameter star formation history of \cite{Johnson2021} instead of the linear-exponential form used in the WAF figure, using a calculational method described in Appendix~\ref{sec:appendix}.  We model the SNIa delay time distribution with a sum of two exponentials that approximates a $t^{-1.1}$ power law \citep{Maoz2017}, with a minimum delay time $t_d=0.15\Gyr$.  For the APOGEE points we use the same sample cuts described in \S\ref{sec:plateau} but restrict to $\Rgal=7-9\kpc$ and $|Z|<2\kpc$, and we downsample the $|Z|<1\kpc$ stars by a factor of four for visual clarity.  As expected, the high-$\alpha$ population is more prominent at high $|Z|$.  

For our ``inner Galaxy'' track we adopt $\eta=0$ (no outflow), a high SFE with $\taustar=0.6\Gyr$, and a star formation history that peaks at $t=2\Gyr$ and declines exponentially ($\taustar=6\Gyr$) at late times.  This track evolves to $\feh\approx 0.3$ and slightly sub-solar $\mgfe$, and it roughly follows the upper envelope of the observed high-$\alpha$ population.  For the ``solar radius'' track we adopt $\eta=0.6$, a lower SFE with $\taustar=2\Gyr$, and a slower rise and shallower decline of the star formation history.  As expected from the discussion in Sections \ref{sec:outflows}-\ref{sec:snia}, this track evolves to $\mgfe\approx\feh\approx 0$.  The ``outer Galaxy'' track adopts $\eta=2$, a low SFE with $\taustar=6\Gyr$, and a SFR that rises slowly and becomes approximately constant.  This track evolves to low $\feh\approx -0.45$, principally because of the strong outflow.

Although Figure~\ref{fig:apogee2} qualitatively resembles the corresponding WAF figure, with lower $\eta$ values compensating for lower yields, the match between the models and the APOGEE data is noticeably worse than in WAF.  The principal reason for this difference is that the lower values of $\eta$ lead to slower growth of metallicity at early times (see Equations~(\ref{eqn:Zevol}) and (\ref{eqn:taubar})), so that the downturn of $\mgfe$ from SNIa enrichment sets in at lower $\feh$.   Only the inner Galaxy track has a knee in $\mgfe$ at roughly the location shown by the APOGEE data. The knee would shift to still lower $\feh$ if we adopted lower SFE or assumed a shorter minimum delay in the $t^{-1.1}$ DTD.  Additionally, the low-$\alpha$ population in this APOGEE DR17 sample exhibits a kinked structure that was not evident in the DR12 sample shown in the WAF figure, and this kink is difficult to obtain with smoothly changing star formation histories. One should not draw sharp conclusions from a simple model overlay like Figure~\ref{fig:apogee2}; for fully realized GCE models with radial mixing see, e.g., \cite{Minchev2013,Minchev2014}, \cite{Loebman2016}, \cite{Johnson2021}, and \cite{Chen2022}.  Nonetheless, Figure~\ref{fig:apogee2} suggests that the low-yield, low-outflow combination favored by the RMN23 measurement makes it more challenging to reproduce observed structure in the $\afe-\feh$ plane.  Other groups have proposed very different origins for the bimodality in $\afe$, such as the two-infall model in which dilution resets the ISM metallicity before the low-$\alpha$ sequence evolves \citep{Chiappini1997,Spitoni2019}, and a clumpy burst scenario in which stochastic local self-enrichment takes regions of low-$\alpha$ ISM temporarily to high-$\alpha$ \citep{Clarke2019,Garver2023}.

Motivated by the $\mgfe-\feh$ trends of {\it in situ} halo stars in the H3 survey, Conroy et al. (\citeyear{Conroy2022}, hereafter C22) propose a different GCE model in which the true $\mgfe$ plateau lies at $+0.6$, and an inflection to rising $\mgfe$ between $\feh\approx -1.5$ and $\feh\approx -0.5$ is caused by a rapidly growing SFE that drives an accelerating star formation rate. \cite{Chen2023} suggest a variant of this model in which the accelerating SFR is driven largely by gas inflow rather than growing SFE.

\begin{figure}
    \centering
    \includegraphics[width=\columnwidth]{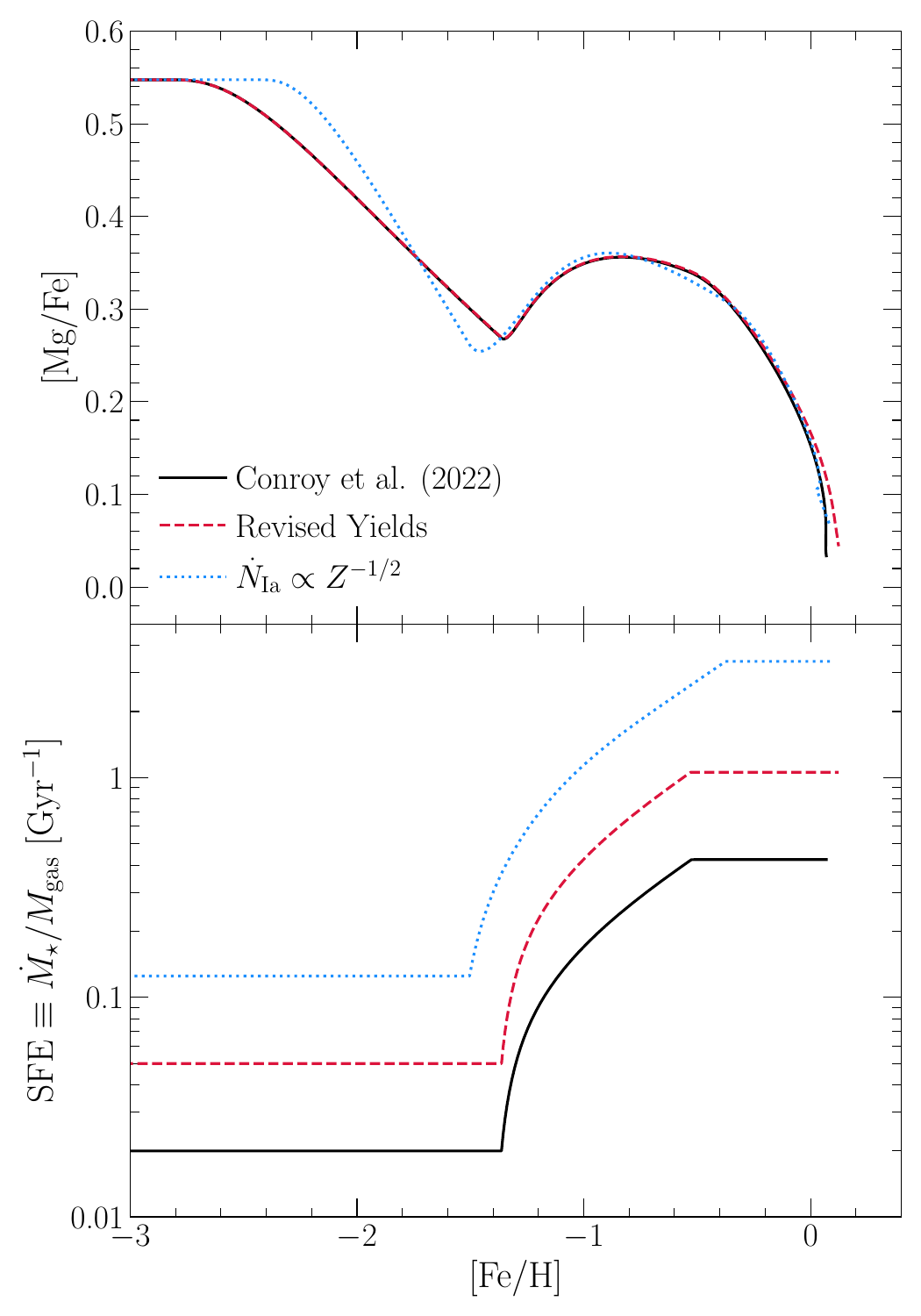}
    \caption{(Top) Evolutionary track of the original C22 model (solid curve), a modified version of this model in which all yields are reduced by a factor of 2.5 and $\eta$ is reduced from 2.0 to 0.3 (dashed curve), and a third version in which the SNIa yield scales with metallicity as $Z^{-0.5}$ for $\mgh>-1$ (dotted curve).  In the modified versions we adjust the time-dependence of SFE to approximately reproduce the $\afe-\feh$ track of the original model.  (Bottom) The SFE ($=\taustar^{-1}$) as a function of $\feh$ for the three models.  All models are computed using the Versatile Integrator for Chemical Evolution ({\tt VICE}, \citealt{Johnson2020}).
    }
    \label{fig:conroy}
\end{figure}

To produce a high plateau and reach $\mgfe \approx 0$ at late times, the C22 model adopts $\ymgcc = 3.45\Zmgsun$ and $\yfecc = 0.875\Zfesun$, and an SNIa Fe yield $\yfeIa = 2.5\yfecc$.  
Figure~\ref{fig:conroy} shows $\afe-\feh$ tracks and the SFE evolution for the original C22 model (solid curves) and a revised version (dashed curves) in which all three yields are reduced by a factor of 2.5, so that $\ymgcc = 1.38\Zmgsun$ matches the value implied by \Eqref{yxccscale}\ for $\mgfecc=0.6$.  We adopt $\eta=0.3$ so that the model evolves to the same metallicity as the original high-yield model, which has $\eta=2$.  We adjust the quantities in Equation (3) of C22 to produce a nearly identical $\afe-\feh$ track, which C22 show to be a good match to the {\it in situ} population measurements for H3 and APOGEE.  

In C22, the $\mgfe$ knee occurs at $\feh \approx -2.75$, which requires an extremely low initial SFE, with $\taustar = 50\Gyr$.  Rapid growth of SFE between 2.5 and $3.7\Gyr$ leads to the bump in $\afe$, followed by another decline when the SFE becomes constant at $t>3.7\Gyr$.  In the revised yield model, the lower $\ymgcc$ and $\yfecc$ must be compensated by a higher initial SFE (shorter $\taustar$) so that $\feh$ reaches $-2.75$ before SNIa enrichment begins.  In fact we find that simply increasing SFE at all times by a factor of 2.5, the same factor by which the yields were reduced, leads to a virtually identical model track once $\eta$ is reduced.  The late-time SFE required to put the {\it second} $\mgfe$ downturn at $\feh \approx -0.6$ is fairly high, with $\taustar=1\Gyr$, though not as high as that of the inner Galaxy curve in Figure~\ref{fig:apogee2} because the yield scale of this $\afecc=0.6$ model is higher.

Based on the observed SNIa rate as a function of galaxy stellar mass \citep{Brown2019}, \cite{Johnson2022snia} argue that the SNIa rate increases with decreasing metallicity, with a scaling similar to the $Z^{-0.5}$ dependence of the close-binary fraction found in APOGEE \citep{Moe2019}.  The dotted curves in Figure~\ref{fig:conroy} show a model in which we set $\yfeIa = 2.5\yfecc(Z_{\rm Mg}/\Zmgsun)^{-0.5}$ for $\mgh> -1$, saturating at $\sqrt{10}$ times the solar metallicity rate.  The high SNIa rate at low metallicity causes a steeper early drop in $\mgfe$, so it is not possible to reproduce the locus of the C22 model exactly.  Nonetheless, we are able to find a model with quite similar behavior by adjusting the SFE history.  This model requires a particularly high SFE at late times so that the second downturn of $\mgfe$ remains at $\feh \approx -0.6$ despite the elevated SNIa rate.  

\section{Conclusions}
\label{sec:conclusions}

We use RMN23's estimate of the mean CCSN Fe yield, $\ybarfecc=0.058\pm 0.007\Msun$, to anchor the scale of population-averaged massive star yields.  Taking $\afecc=0.45$, a Kroupa IMF, and a massive star explosion fraction $\Fexp=0.75$, we find that the population-averaged dimensionless yields $\yocc$ and $\ymgcc$ are nearly equal to their corresponding solar mass fractions (\Eqref{yxccscale}).  In other words, for every solar mass of stars formed, on average, massive stars release a mass of freshly synthesized O and Mg equal to the mass of these elements in the sun.  For Si we estimate $\ysicc \approx 0.8\Zsisun$ based on the lower observed [Si/Fe] plateau and the expectation that a fraction of solar Si comes from SNIa.  Our estimated ``$1\sigma$'' uncertainty in $y_{\alpha}^{\rm cc}/Z_{\alpha,{\odot}}$ is about 0.1 dex ($\sim$ 25\%), contributed by the uncertainties in $\afecc$, in $\Fexp$, and in the RMN23 measurement itself.

Models with a Kroupa IMF in which all $M>8\Msun$ stars (or even all $M=8-40\Msun$ stars) explode as CCSN predict a $\yocc$ that is $2-3$ times higher than our empirically inferred value (see, e.g., \citealt{Andrews2017}, G21).  The lower yield scale reduces the need for outflows in GCE models to reproduce observed abundances, though for our fiducial choices we still find that $\eta = \Mdotout/\Mdotstar \approx 0.6$ is required to produce a solar ISM abundance at late times.  Many models of the galaxy mass-metallicity relation have assumed a higher yield scale. Adopting our empirical $\yocc$ would reduce the $\eta$ values inferred in these studies, by about a factor of two in the low mass systems with $\eta \gg 1$.  GCE models with our empirical $\yocc$ predict an ISM D/H ratio that is 71\% of the primordial D/H (\Eqref{xd}), which is lower than the ISM D/H estimated by \cite{Linsky2006} but consistent at the $2\sigma$ level.  Improved determination of the D/H ratio in the local ISM --- or even better, at locations that probe a range of ISM metallicities --- would provide a powerful consistency test for our understanding of stellar yields and chemical evolution.

Assuming a Kroupa IMF, the RMN23 value of $\ybarfecc$ and our inferred $\yocc$ are well reproduced by the S16 CCSN models with the Z9.6+N20 neutrino-driven engine, which corresponds to an explodability threshold $e_0=0.046$ in the terminology of G21.  Thus, these empirical yields can be explained by massive star nucleosynthesis calculations with a plausible level of black hole formation.  However, with this black hole formation landscape the S16 models underpredict the empirical yields of Mg and Si by factors of 3.5 and 2.5, a discrepancy already evident in their underprediction of the solar Mg/O and Si/O ratios.

By requiring that GCE models approach $\afe \approx 0$ at late times, we can estimate the population-averaged SNIa Fe yield $\yfeIa$, again with a dependence on the assumed $\afecc$ (\Eqref{yfeIascale}).  Assuming $\afecc=0.45$ and a mean Fe yield per SNIa of $0.7\Msun$ \citep{Howell2009}, we infer a Hubble-time integrated SNIa rate $\RIa = 1.1 \times 10^{-3}\,\Msun^{-1}$, consistent with the normalization $(1.3 \pm 0.2)\times 10^{-3} \,\Msun^{-1}$ that \cite{Maoz2017} found by comparing the cosmic histories of star formation and SNIa.  Our estimate assumes that $\RIa$ and $\yfeIa$ are metallicity independent.  If the SNIa rate depends on metallicity \citep{Johnson2022snia} then more careful modeling is needed to relate the CCSN and SNIa yield scales.

Theoretical uncertainties in massive star yields, especially those associated with black hole formation, make {\it ab initio} predictions of the population-averaged $\alpha$-element yield uncertain at the factor-of-three level.  The degeneracy between yields and outflows, on the other hand, makes it difficult to infer the absolute scale of yields from GCE modeling of observed stellar and ISM abundances.  The measurement of $\ybarfecc$ by RMN23 is arguably the best empirical anchor for the scale of stellar yields that is currently available.  Analogous measurements for other elements might be possible with careful light-curve, spectroscopic, or supernova remnant analyses of samples that span the full range of the CCSN population.  The scale of yields has wide-ranging implications for massive star models, galaxy evolution, and the cosmic metal budget.  The central role of yields highlights the importance of stress-testing the $\ybarfecc$ measurement and sharpening its precision, and of improving the determination of $\afecc$ through accurate measurements in low metallicity stellar populations that span a range of galactic environments.

\section{Acknowledgements} 

We thank Dan Maoz for helpful comments on a draft version of this manuscript.  We are grateful to him and to 
Brett Andrews,
Charlie Conroy,
Jennifer Johnson,
Chris Kochanek,
Molly Peeples,
Ralph Schoenrich,
Krzysztof Stanek,
Tuguldur Sukhbold, and
Michael Tucker 
for valuable conversations on these topics over many years.
This work is supported by NSF grants AST-1909841 and AST-2307621.
E.J.G. is supported by an NSF Astronomy and Astrophysics Postdoctoral Fellowship under award AST-2202135.

\appendix
\section{Analytic GCE for a 2-parameter SFH}
\label{sec:appendix}

WAF derive analytic solutions for the evolution of $\alpha$-element and Fe abundances in a one-zone GCE model.  These solutions assume time-independent values of $\eta$ and $\taustar$, metallicity-independent yields, instantaneous CCSN enrichment, and SNIa enrichment with an exponential DTD.  A sum of two exponentials can be used to approximate the $t^{-1.1}$ power-law DTD advocated by \cite{Maoz2017}.  WAF present solutions for a constant SFR, an exponential SFH with $\Mdotstar \propto e^{-t/\tausfh}$, and a linear-exponential SFH with $\Mdotstar \propto t e^{-t/\tausfh}$.  The $\alpha$-element (pure CCSN) solution for an exponential SFH is \Eqref{Zevol} of this paper.

The linear-exponential SFH (also known as ``delayed tau'') has gained popularity as a one-parameter model that captures the rise-and-fall behavior typical of galaxies in semi-analytic calculations and cosmological simulations (e.g., \citealt{Lee2010,Simha2014,Carnall2019}).  However, in this model the rise to the peak and the subsequent exponential decay are tied to each other --- one cannot have a fast rise and slow decline, or vice versa.  The 2-parameter SFH used by \cite{Johnson2021},
\begin{equation}
    \Mdotstar(t) = K \left(1-e^{-t/\tau_1}\right) e^{-t/\tau_2}~,
    \label{eqn:risefall}
\end{equation}
is much more flexible.  For $\tau_1 \ll \tau_2$, the SFH rises linearly at early times, reaches a maximum at $t \sim \tau_1$, then declines exponentially with a timescale $\tau_2$.  Limiting cases include a pure exponential SFH ($\tau_1=0$), a constant SFR    ($\tau_1=0$, $\tau_2\rightarrow\infty$) and a linearly rising SFH ($\tau_1\gg t$, $\tau_2 \rightarrow \infty$).  The normalization constant $K$ scales the overall stellar mass of the galaxy but cancels out in the evolution of chemical abundances.  A GCE model with constant SFE in which the gas supply starts at zero and the gas {\it infall} rate is $\propto e^{-t/\tau_{\rm inf}}$ (e.g., \citealt{Spitoni2017}) follows the SFH of \Eqref{risefall} with $\tau_2=\tau_{\rm inf}$ and
\begin{equation}
    \tau_1 = {\taustar \over 1+\eta-r-\taustar/\tau_{\rm inf}}
    \label{eqn:tau1}
\end{equation}
(see Equation (129) of WAF).

Fortunately, the analytic solutions derived by WAF can be readily applied to this 2-parameter ``rise-fall'' SFH.  To see how, it is useful to write
\begin{equation}
    K\left(1-e^{-t/\tau_1}\right) e^{-t/\tau_2} = K e^{-t/\tau_2} - K e^{-t/\tau_h}
    \label{eqn:risefall2}
\end{equation}
with
\begin{equation}
    \tau_h = \left(\frac{1}{\tau_1}+\frac{1}{\tau_2}\right)^{-1}~.
    \label{eqn:tauh}
\end{equation}
Equations~(50) and~(53) of WAF give solutions for the ISM mass fraction $Z(t)$ of, respectively, a pure CCSN element and the SNIa contribution to Fe, for an exponential SFH.  Equation~(117) of WAF shows how to combine the solutions for any two star formation histories to obtain a solution for the SFH that is the sum of these histories.  In this case, we simply take the solutions for exponential histories with timescales $\tau_2$ and $\tau_h$ and combine them using this equation.  The second exponential SFH has a negative normalization, but the overall SFH is positive-definite so this does not cause unphysical results.  We have confirmed that this procedure reproduces the results of direct numerical integrations, as expected.

The [Mg/Fe] vs. [Fe/H] tracks in Figure~\ref{fig:apogee} are computed using this analytic method.  The three SFHs shown correspond to $(\tau_1,\tau_2) = (1,6)$ for the inner Galaxy track, (2,12) for the solar radius track, and (4,30) for the outer Galaxy track, with all times in Gyr.

\bibliography{yield}{}
\bibliographystyle{aasjournal}

\end{document}